\documentclass[fleqn,usenatbib]{mnras}

\usepackage{newtxtext,newtxmath}
\usepackage[T1]{fontenc}
\usepackage{ae,aecompl}

\pdfoutput=1
\usepackage{graphicx}	% Including figure files
\usepackage{amsmath}	% Advanced maths commands
\usepackage{wasysym}    % Solar symbol, etc
\usepackage[dvipsnames]{xcolor} % To get a wider range of font colours
\usepackage[normalem]{ulem} % To strike out text

\usepackage{etoolbox}

\definecolor{grey}{gray}{0.75}
\definecolor{darkgreen}{rgb}{0.0,0.75,0.0}

\newcommand{\eg}[0]{$\textnormal{e.g. }$}
\newcommand{\ie}[0]{$\textnormal{i.e. }$}
\newcommand{\tn}[1]{\textnormal{#1}}
\newcommand{\sub}[1]{_{\textnormal{#1}}}

\title[Gas flows in Auriga discs]{The parametrization of gas flows in discs in the Auriga simulations}

\author[P. Okalidis et al.]{Periklis Okalidis,$^{1}$\thanks{okalidis@mpa-garching.mpg.de}
Robert J. J. Grand,$^{1}$
Robert M. Yates,$^{1,2}$
Guinevere Kauffmann$^{1}$
\\
$^{1}$Max-Planck-Institut f{\"u}r Astrophysik, D-85741 Garching, Germany\\
$^{2}$Department of Physics, University of Surrey, Surrey GU2 7XH, United Kingdom
}

\date{Accepted XXX. Received YYY; in original form ZZZ}

\pubyear{2020}

\begin{document}
\label{firstpage}
\pagerange{\pageref{firstpage}--\pageref{lastpage}}
\maketitle

\begin{abstract}

We study the radial motions of cold, star-forming gas in the secular evolution phase of a set of 14 magnetohydrodynamical cosmological zoom-in simulations of Milky Way-mass galaxies. We study the radial transport of material within the disc plane in a series of concentric rings. For the gas in each ring at a given time we compute two quantities as a function of time and radius: 1) the radial bulk flow of the gas; and 2) the radial spread of the gas relative to the bulk flow. Averaging the data from all the halos, we find that the radial spread increases with radius in the form of a power law with strong secondary dependencies on the fraction of accreted material and the local radial velocity dispersion of the gas. We find that the bulk motion of gas is well described in the inner disc regions by a radially-independent mean inward flow speed of $-2.4\,\tn{km s}^{-1}$. The spread around this value relates to the change in angular momentum of the gas and also the amount of accreted material. These scalings from fully cosmological, MHD simulations of galaxy formation can then be used in semi-analytic models to better parameterise the radial flow of gas in discs.
%\robyc{don't need the minus sign here}
\end{abstract}

\begin{keywords}
galaxies: evolution -- galaxies: disc -- galaxies: kinematics and dynamics -- methods: data analysis
\end{keywords}

%%%%%%%%%%%%%%%%%%%%%%%%%%%%%%%%%%%%%%%%%%%%%%%%%%

%%%%%%%%%%%%%%%%% BODY OF PAPER %%%%%%%%%%%%%%%%%%
\color{black}
\section{Introduction}

Examining the kinematics and flow of gas within the disc can give us useful insights into some key aspects of the disc evolution. Gas inflowing through the disc plane is directed to the central regions, fuelling star formation. Furthermore, radial flows result in mixing of metal poor gas accreted in the outer regions of the disc with more metal-enriched gas due to the stellar evolution in the plane and can influence the metallicity gradients we observe in disc galaxies \citep{Spitoni11,Schoenrich17,Yates20b}. Similarly, the redistribution of gas due to these flows determines the locations of star formation hence influences the star-formation rate (SFR), stellar and gas density profiles.

There have been studies of, and recent interest in, how gas flows across the virial radius of dark matter (DM) haloes (\eg{}\citealt{Nelson15}) and eventually reaches the central galaxy. However, in the field of numerical simulations there are relatively fewer studies concerning how gas flows in the plane of the disc affect the galaxies within these haloes. 

Therefore, in this study, we focus on the path of the gas inside the galactic disc. The gas that is in place in the disc along with the newly accreted gas \citep{Stevens17}, are subject to angular momentum loses, resulting in infalls towards the centre of the potential well, while following the rotational pattern of the galaxy. The collisional nature of the gas means that turbulent behaviour can become important, while the gas is also subject to external torques from surrounding subhalos or non-axisymmetric structures such as bars. 

Radial gas flows have been studied in early work by \cite{Lacey85}, who concluded that flows of the order of a few ${\rm km s^{-1}}$ are necessary in their galactic disc models to reproduce the exponential gas density profiles observed in discs \citep{Bigiel12,Wang14}. Their arguments for the emergence of radial flows were based on physical grounds relating to three processes. Firstly, the viscosity of the gas whereby the gas clouds interact which each other, dissipating energy and leading to inwards flows. Secondly, the angular momentum difference between the newly accreted onfalling gas and the gas already present in the disc. And thirdly, the presence of non-axisymmetric density patterns, such as bars and spirals arms, which can add or remove angular momentum from the gas. 

Following this work, many models that study the evolution of disc galaxies include recipes for the transport of gas mass within the disc, usually by modelling the fluxes across different radii, or the radial inflow velocity of gas at a given radius \citep{Kubryk15a,Cavichia14,Bilitewski12,Schoenrich09}. These recipes, based on the physical grounds laid-out by \cite{Lacey85}, are necessary in most cases to reproduce the observed metallicity profiles and construct accurate chemical evolution models. 

From a theoretical perspective, \cite{Krumholz18} have developed a model that includes radial transport of gas via differential equations which depends on parameters such as the surface density and velocity dispersion of the gas, the presence of non-axisymmetric torques and also energy injection and dissipation from star-formation feedback and turbulence. This model is based on previous works \citep{Krumholz10,Forbes12,Forbes14} that were aimed at establishing the processes that relate to the radial mass transport in discs. These developments are very useful in constructing advanced semi-analytic models that include radial transport of gas and stars \citep{Forbes19}. Similarly, \cite{Stevens18}, using the \textsc{DARK SAGE} semi-anlytic model \citep{Stevens16}, allows for radial transport of material in the discs, transferring mass between different annuli when there is a gravitational instability in a given annulus, while conserving the angular momentum in the process.

From an observational perspective, gas movement in the disc plane can be studied using high-resolution 21 cm atomic hydrogen (HI) \citep{Sellwood10,Schmidt16,Speights19} or CO \citep{Wong04} gas maps of nearby galaxies. These studies look for residual non-circular components of the gas motions in the disc by removing the bulk rotational motions. They consistently report radial speeds in the range of a few km s$^{-1}$ towards the center (\ie{}inflows). \cite{Schmidt16} have found evidence of inflowing gas in most of the HI THINGS galaxy sample, but also find some galaxies with no clear inflow, and some with outward gas motions or more complex kinematics, showing that there is substantial variation between different galaxies. 

Using zoom-in simulations of disc galaxies, \cite{Nuza19} have measured fluxes for the gas through cylindrical shells at given radii, looking separately at the inflowing/outflowing gas but also for the fluxes of gas leaving/entering the disc in the perpendicular direction. They report net inwards radial flux in the discs, which is more pronounced in the inner regions and also during the presence of merger events. \cite{Goldbaum15,Goldbaum15b} have run isolated disc simulations with and without star formation feedback to study the effect of gravitational instability driven turbulence as a mass transport mechanism in discs. They conclude that the gravitational instability, expressed by the Toomre Q parameter, is a dominant source of radial transport of material even when feedback is present and they find that this transport of gas is sufficient to fuel the star formation in the inner part of discs. They show radial profiles of gas mass fluxes in the disc, measuring fluxes of the order of $\sim 1 M_{\odot}$ yr$^{-1}$ with high variability around the median values at any given radius, with both radially inwards and outwards flows dominating at different times. 

With the advent of new generations of high-resolution simulations and numerical codes, modelling gas flows has become more detailed and accurate. Many simulations have also managed to reproduce disc-dominated, rotationally-supported, star-forming systems (\eg{}\citealt{Font20,Marinacci14,Aumer13,Agertz13}) and have also studied bar formation \citep{Fragkoudi20}. Driven by these advances, we are opting to use the Auriga simulation suite \citep{GGM17} as a means to study detailed gas flows in galactic discs. The gas properties in the Auriga simulations have been studied in \cite{Marinacci17}, finding good agreement with observed properties such as the extent of the gas disc and the radial gas profiles. It has been established in many simulations that merger events are drivers of gas flows to the central regions of galaxies \citep{Bustamante18}. Furthermore, bars have been shown to be responsible for strong gas flows within the co-rotation radius. In this study, we focus more on the epochs of the disc galaxies evolution that are free of major merger events, in order to examine the gas inflow that arises from the internal processes of the disc evolution or smooth gas accretion from the environment.

Our approach is to use our knowledge of gas flows gained from the Auriga simulation to provide parametrisations that can be readily implemented into semi-analytic models (SAMs) of galaxy formation. More specifically, we would like to later apply the results of this study to the \textsc{L-Galaxies} SAM, that has recently been updated to include radial rings that allow the study of radial dependencies in galactic discs \citep{Henriques20}. The new model version also includes the radial flow recipe presented by \cite{Fu13}, which allows gas to be transferred from outer to inner rings with an inflow speed proportional to the galactocentric radius of the gas. SAMs have the advantage over hydrodynamical simulations of requiring shorter computational times, allowing for an easier exploration of the parameter space describing sub-grid physical processes, and thus helping us understand which processes are primary and which are secondary in influencing different observational phenomena.

We structure this paper as follows. First, we outline the Auriga galaxy formation model and the characteristics of the halos that we choose to use. Then, we describe our analysis, which is done using the tracer particles that are implemented in the Auriga runs and is based on a decomposition of the galactic discs into a set of concentric radial rings. In the next section we present our results, looking at the effect of several physical quantities on the process of radial gas inflow and finally, we extract parametrisations that describe this process and we provide a basic method for including these in the context of a semi-analytic model. 

\section{Simulations}

Auriga is a set of high resolution, magneto-hydrodynamical cosmological ``zoom'' simulations for the formation of Milky-Way-mass galaxies. Our sample for this study comprises 14 Auriga halos; 6 halos from the original simulation suite \citep{GGM17} with a halo mass\footnote{Defined to be the mass inside a sphere in which the mean matter density is 200 times the critical density, $\rho _{\rm crit} = 3H^2(z)/(8 \pi G)$.} in the range $1-2 \times 10^{12} M_{\odot}$, and 8 simulations of slightly lower halo masses of $0.5-1 \times 10^{12} M_{\odot}$ \citep{GVZ19}. We have selected these halos because they include tracers particles which are necessary for our analysis. In addition to their mass, halos are selected based on a mild isolation criterion from the $z=0$ snapshot of the dark matter-only counterpart to the cosmological Eagle simulation of comoving side length 100 cMpc (L100N1504) introduced in \citet{SCB15}. The cosmological parameters that are used are $\Omega_{m}=0.307$, $\Omega_{b}=0.048$, $\Omega_{\Lambda}=0.693$, $H_{0}=100h$ $\rm km\, s^{-1}\, Mpc^{-1}$ and $h=0.667$, taken from \citet{PC13}.

The initial conditions of the zoom simulations are set at $z=127$. The high-resolution regions of these simulations have a mass resolution of $\sim 5\times10^{4}$ $\rm M_{\odot}$ per baryonic element and a comoving softening length of 500 $ \rm pc$ $h^{-1}$. The physical softening length grows until $z=1$, after which time it is kept fixed. The physical softening value for the gas cells is scaled by the gas cell radius (assuming a spherical cell shape given the volume), with a minimum softening set to that of the collisionless particles.

The simulations are then evolved forward in time with the quasi-Lagrangian magneto-hydrodynamics code \textsc{arepo} \citep{Sp10,PSB15} and a galaxy formation model that includes the physical processes important for the formation and evolution of galaxies \citep[for a detailed overview, see][]{GGM17}. In \textsc{arepo}, gas cells are modelled with an unstructured mesh in which gas cells move with the local bulk flow. The galaxy formation model includes primordial and metal-line cooling \citep{VGS13} and a prescription for a spatially uniform background UV field for reionization. Gas that becomes denser than $0.11$ atoms $\rm cm^{-3}$ is considered part of the star-forming interstellar medium (ISM), which is modelled as a two phase medium: cold clouds embedded in a hot, volume filling phase \citep{SH03} assumed to be in pressure equilibrium. Star particles form stochastically from this gas following a Schmidt-type star formation law, and are modelled as Simple Stellar Populations (SSPs) defined by an age, mass and metallicity. The stellar evolution model follows type Ia supernovae (SNe-Ia) and winds from Asymptotic Giant Branch (AGB) stars that return mass and metals (9 elements are tracked: H, He, C, O, N, Ne, Mg, Si and Fe) to the surrounding gas. Supernovae type II are also assumed to return mass and metals following the instantaneous recycling approximation. Galactic winds from SNII are modelled by the wind particle scheme for non-local energetic feedback \citep{VGS13}, which effectively models the removal of mass from star-forming regions and deposits mass, momentum and energy into gas of density lower than $5\%$ of the density of star-forming gas. The model includes prescriptions for the accretion of gas onto black holes and energetic feedback from Active Galactic Nuclei \citep[as described in][]{GGM17}. Magnetic fields are seeded at $z=127$ with a comoving field strength of $10^{-14}$ cG \citep{PMS14}. The magnetic field strength in the Milky Way-like halo has been shown to quickly amplify to a strength and radial profile in excellent agreement with observations \citep{PGG17,PGP18,PVB19}.

For each halo we have 252 snapshots down to redshift 0, spaced at intervals ranging between 45 - 75 Myr with a median value of $\sim 60$ Myr.

\begin{figure*}
    \centering
    \includegraphics[width=\textwidth]{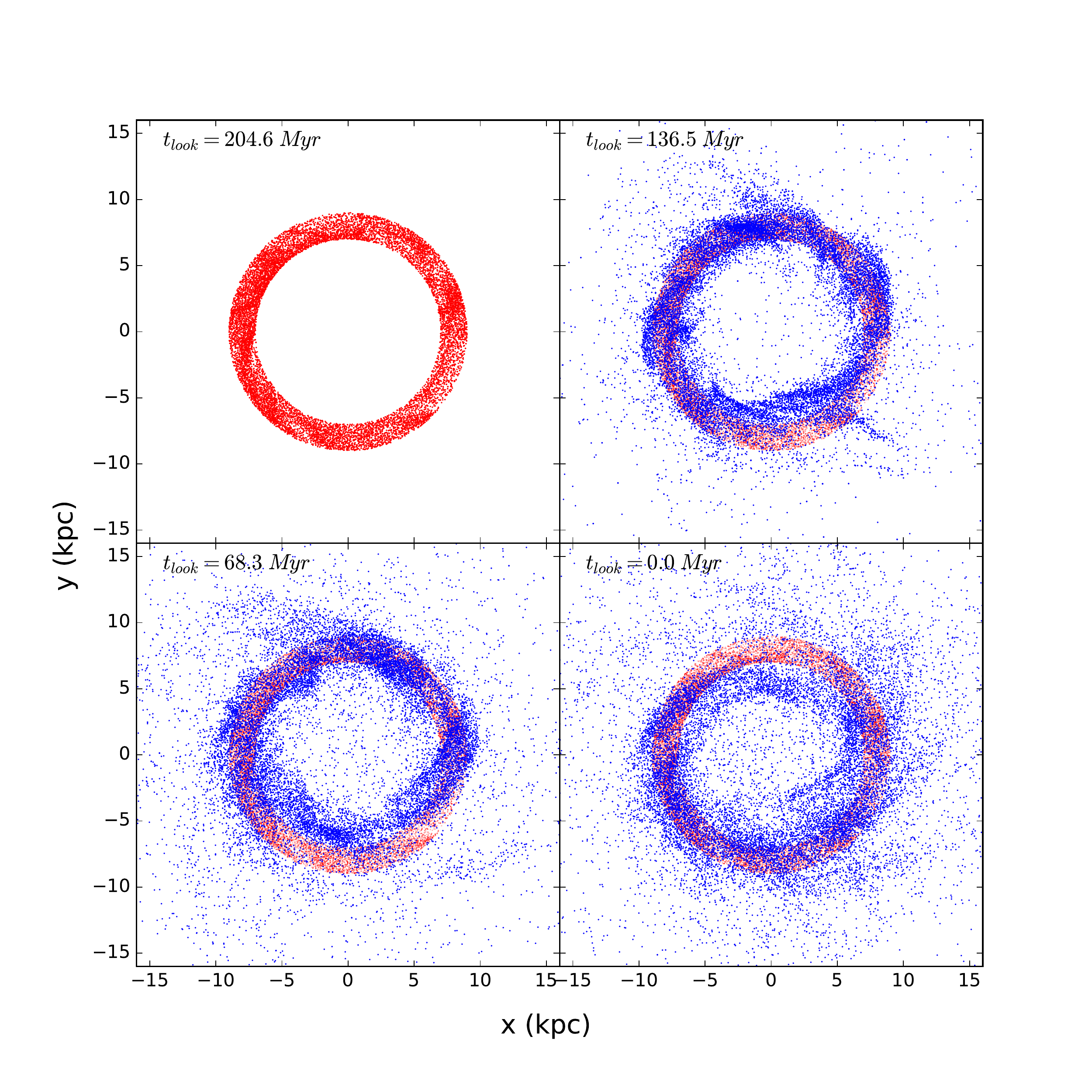}
    \caption{Top left: Tracers selected in a ring centered at 8kpc for one of the haloes. Shown here are their x and y positions in the plane of the disc. The next three panels show the evolution of the planar distribution of tracers at the next three snapshots, with the lookback time from redshift 0 quoted on top.}
    \label{mvFig}
\end{figure*}

\section{Methods}

\subsection{Tracer particles}
Owing to the quasi-Lagrangian nature of the \textsc{arepo} code, gas cells move both with the bulk local gas flow and advect mass across their boundaries to neighbouring cells. In order to track the evolution of fluid elements, therefore, we need to follow tracer particles that connect gas cells at different snapshots in time. The tracers are initialized at the beginning of the simulation with one tracer particle per gas cell. Tracers can move across neighbouring cell faces in a probabilistic way depending on the ratio of the outward-moving mass flux across the face and the mass of the cell, which is essentially a Monte Carlo sampling of the outward mass flux for each gas cell in the simulation box \citep{Genel13,GVZ19,DeFelippis17}. 

Tracer particles are not exclusively locked in the gas state but can occupy five different cell/particle types depending on the physical processes they are subject to: 
\begin{itemize}
    \item non-star forming gas cells
    \item star-forming gas cells
    \item wind particles
    \item star particles    
    \item black hole particles
\end{itemize}

A tracer can alternate between the different states. For example if a star-forming gas cell creates a new star particle, the tracer associated initially with the star-forming gas cell will subsequently track the star particle. Tracers can also alternate between the star-forming (SF) and non-star-forming (non-SF) gas phases based on their cell density. Thermal dumps from AGN feedback can directly heat SF to non-SF gas, while cooling processes naturally change non-SF gas to SF. In addition, tracers can transfer into wind particles via supernova activity and potentially return via fountain flows at a later time \citep{GVZ19}. Finally, tracers can move from star particles back to gas cells via stellar evolution, \eg{}AGB winds, though this is not a dominant pathway, as \cite{GVZ19} find that comparatively small number of tracers move from star particles to gas cells via AGB winds compared to supernova events.

The Auriga simulation volume is a cube of side length equal to 100 Mpc, with the high-resolution region around the central galaxy being of order 1 Mpc (no low-resolution particles/gas cells are found within this region). In this project, we are interested in the kinematics of the main disc galaxy which in the majority of cases is under $50$ kpc in diameter with regard to both its stellar and gas content. Matter structures farther than a few times the disc radius at any given snapshot should not immediately influence the gas flows in the disc, however they may become relevant at a subsequent snapshot. For example, a subhalo just entering the virial radius of the main halo does not influence the central disc. However the material (hence the tracers) carried by this subhalo may potentially become part of the main disc at a later time, should it merge with the main galaxy. Tracers locked in structures that never arrive at the vicinity of the main galaxy are thus ignored during the analysis. 

We make a selection of all the tracers which at the final snapshot of the simulation are within a radius of 500 kpc from the centre of the galaxy. When initially selecting tracers, we do not differentiate between those in the gas phase, winds or in stars, since a tracer locked in a star particle at $z=0$ was most likely in the gas phase at an earlier time and hence was part of the gas inflow that we study. The gas tracers at the final snapshot that are inside or in the vicinity of the disc could either have been in place from early times or been accreted at a later stage smoothly or by merging. Our radial cut is sufficiently large that tracers are unlikely to escape this boundary even if they are launched in winds, ensuring that we do not lose information about the flow elements even at earlier times. Once selected, tracers can be tracked back in time to get information on their positions and velocities.

\begin{figure*}
\centering
\begin{minipage}{.5\textwidth}
  \centering
  \includegraphics[width=\linewidth]{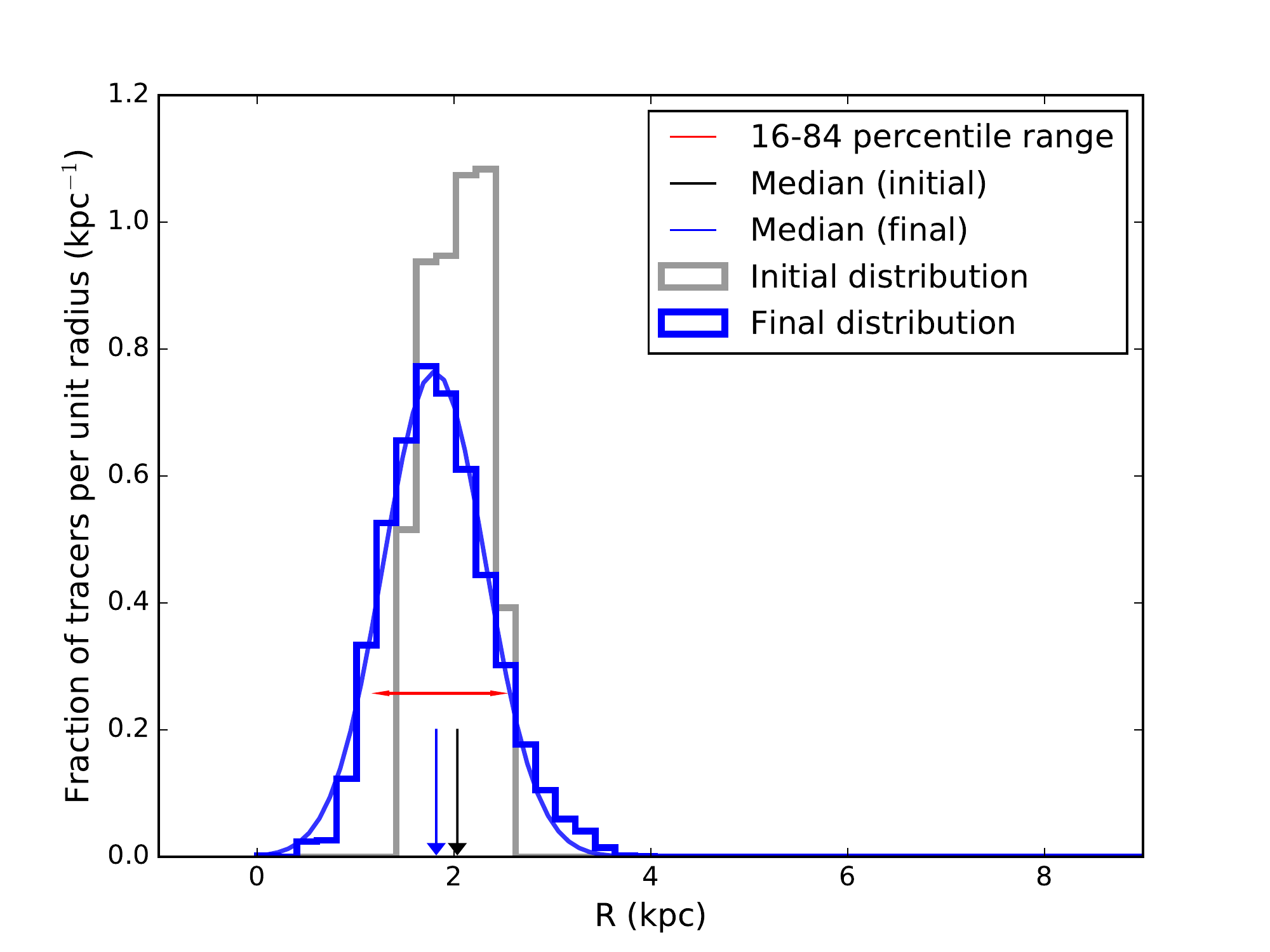}
\end{minipage}%
\begin{minipage}{.5\textwidth}
  \centering
  \includegraphics[width=\linewidth]{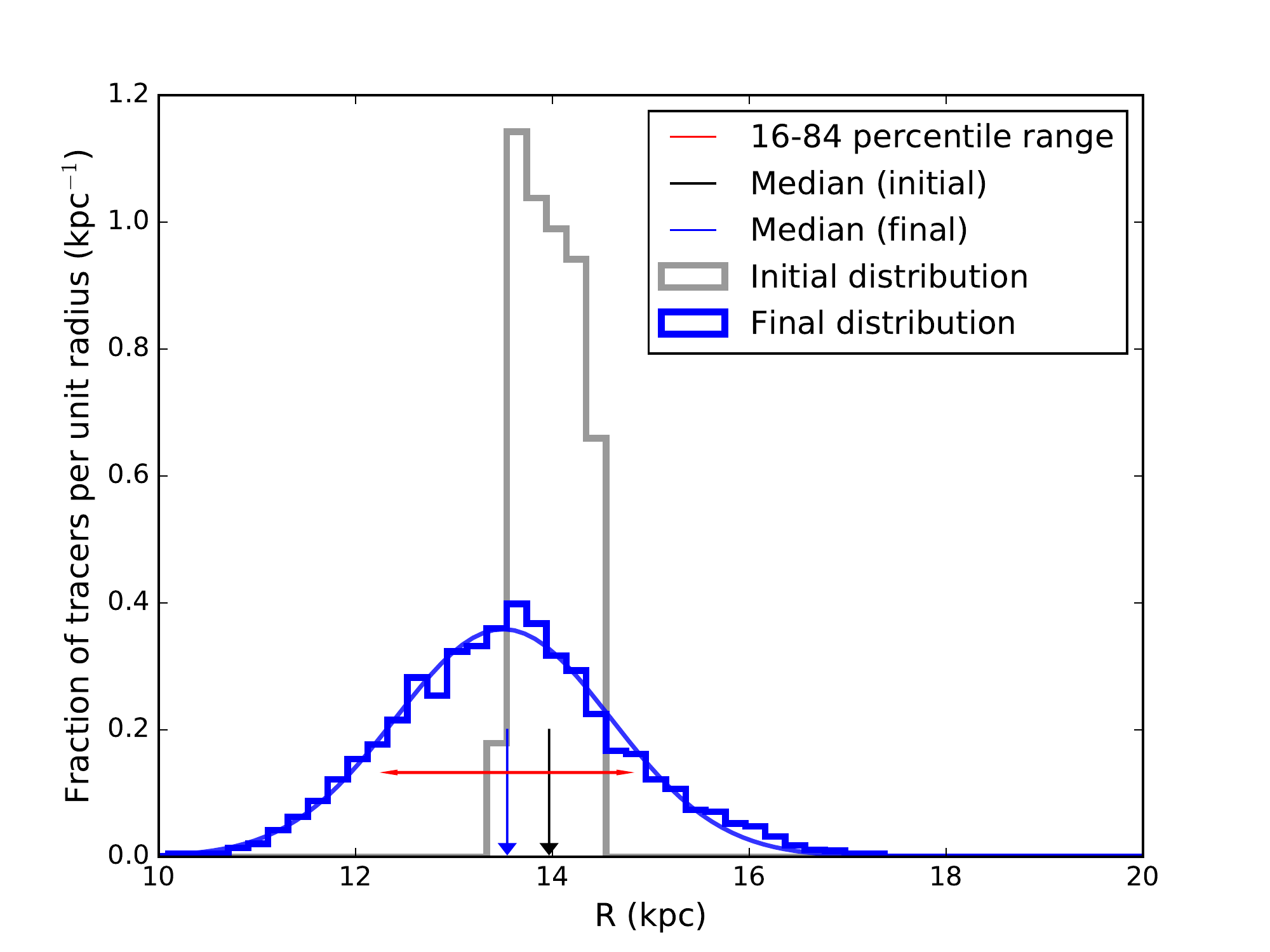}
\end{minipage}
\begin{minipage}{.5\textwidth}
  \centering
  \includegraphics[width=\linewidth]{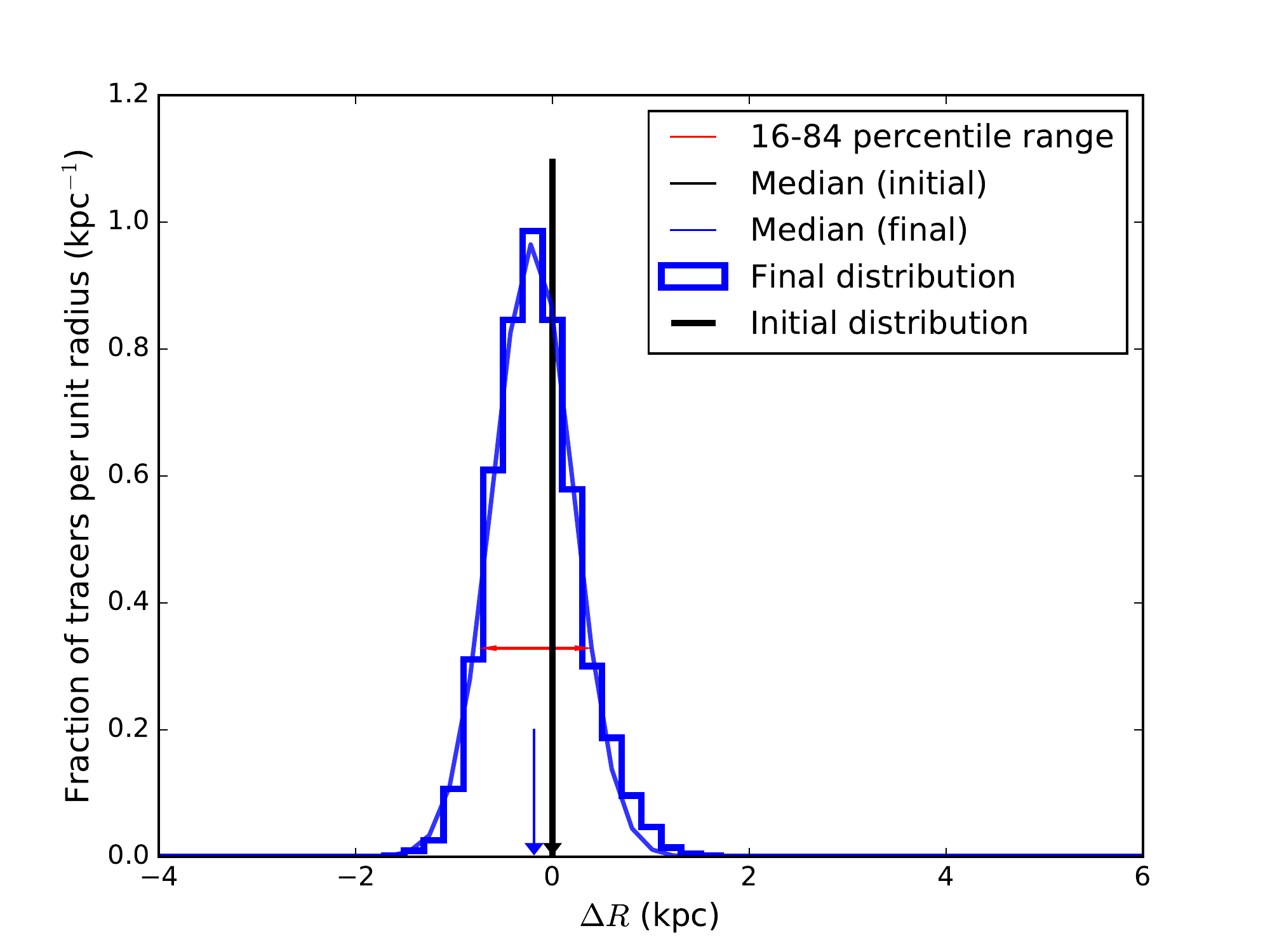}
\end{minipage}%
\begin{minipage}{.5\textwidth}
  \centering
  \includegraphics[width=\linewidth]{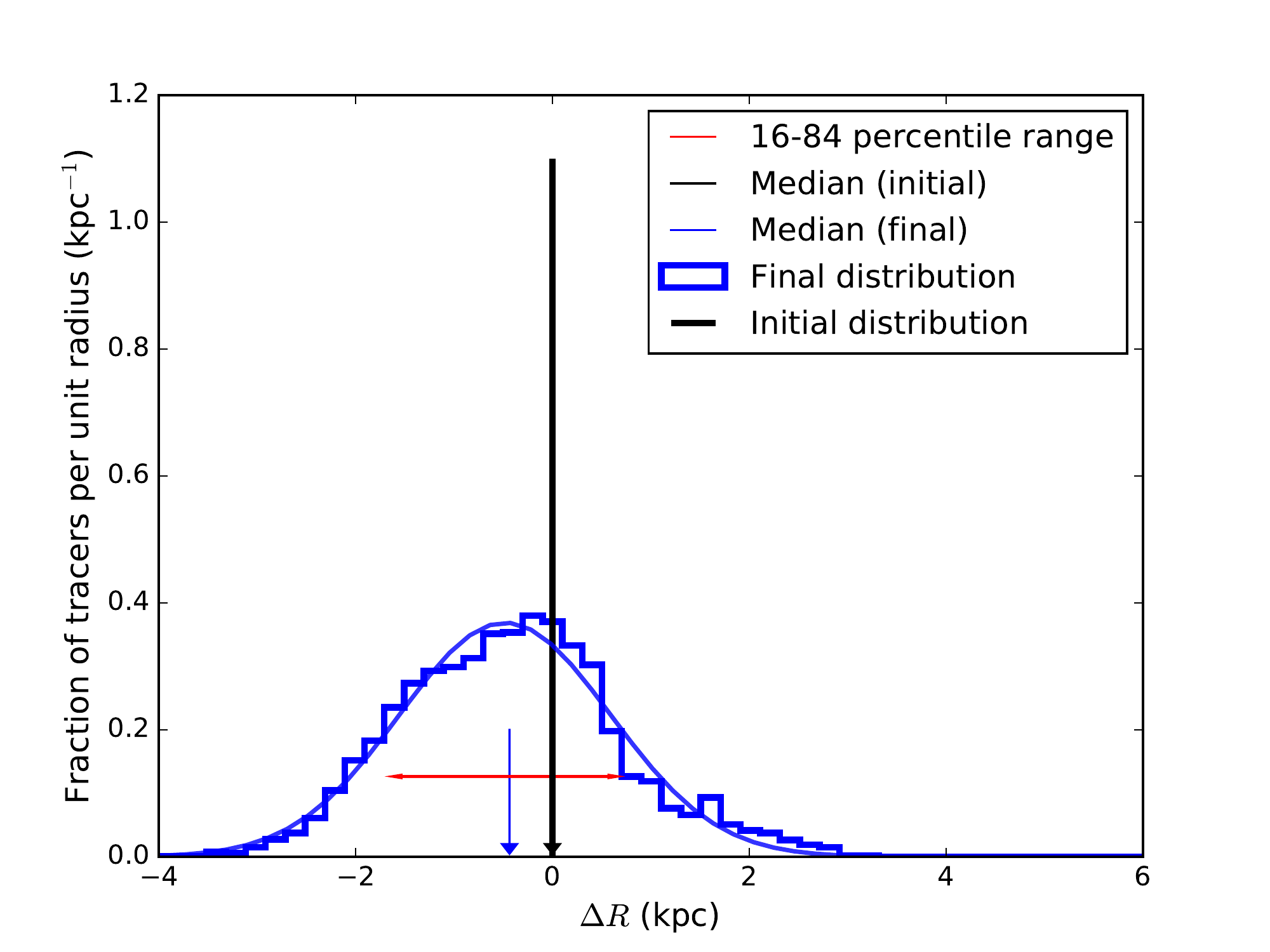}
\end{minipage}
\caption{Examples of the histograms that are computed in order to extract the information for the dispersion of tracers. At the top row we show the histograms in terms of the galactocentric radius of the tracers. The initial distributions (in gray) are calculated at snapshot n and the final (in blue) at the subsequent snapshot n+1. At the bottom row we show the histograms for the same rings in terms of the difference in galactocentric radii $\Delta R = R_{n+1;n} - R_{n}$. In this case the distribution at snapshot $n$ is a delta function whereas the one at $n+1$ displays the spread and mdedian shift of the tracers. The median of the distributions in all cases is marked with the vertical arrows. Over-plotted is the Gaussian fit to the final distribution. The red horizontal arrows show the 16-84 percentile ranges of the final distribution. Here we select the cases for an inner ring (left) at $\sim2$ kpc and an outer ring (right) at $\sim14$ kpc to show the difference in the spread of the tracers.}
\label{tracerhists}
\end{figure*}

\subsection{Ring analysis}

Motivated by the implementation of ring decomposition of the cold gas disc in \textsc{L-Galaxies}, we decide to perform a similar kind of ring analysis in Auriga. We aim to have a description of the kinematics of the gas that belongs to a ring centred at a given galactocentric radius in the plane of the disc.
We split galactic discs into a series of concentric rings of equal width extending out to 20 kpc from the galactic centre and 2 kpc above and below the galactic plane. In order to do this, we rotate the coordinate system of the simulation box so that the plane of the disc is described by the x and y coordinates, and the z-coordinate indicates the distance above or below the plane. The disc plane itself is determined using information on the angular momentum of the stellar disc in the simulation \citep[as described in][]{GGM17}. More specifically, the z-axis of the disc plane is identified by calculating the dot product of the eigenvectors of the moment of inertia tensor of star particles within 0.1 $R_{200}$, with the angular momentum vectors of the same star particles in the coordinate reference frame of the simulation box. The eigenvector of the inertia tensor that is most closely aligned with the principal angular momentum axis is chosen as the z-axis.

Our height and radius cuts are chosen so that they include most of the cold gas that comprises the disc in the majority of cases. The radial cut was selected by inspecting the extent of the cold gas distribution in the different halos. In only one halo did the cold gas disc extend further than 20 kpc, but for the rest of the cases the disc was fully included within the cut. Cold gas tracers above our height cut are not directly associated with the radial motions in the disc that we want to study, but are rather in an accretion phase perpendicular to the disc plane. These tracers are also at a much lower density, so would not significantly contribute to the median properties of the flows we compute for the disc.

Each ring is simply characterised by its galactocentric radius. Given that the extent of cold gaseous discs varies between haloes and snapshots, we choose to normalise the radius of each ring by dividing by the disc radius, of the star-forming gas disc at each snapshot. The radius is calculated as the radius which encloses 95 per cent of the star forming gas in the disc, hence we name it $R_{95}$. We choose this definition for the disc edge, instead of 100 per cent of the SF gas, to account for cases where blobs of cold gas are potentially accreting in the outer edges of the disc without yet constituting part of it. It should be noted that we do not vary the width of the rings between galaxies or snapshots, and we also use the same number of rings (20) in each case.  

For each ring, we identify the tracers in the star-forming gas phase that lie in it at snapshot $n$ and then ask what the positions of these tracers are at the next snapshot $n+1$. 
In the absence of major disturbances in the disc, a given parcel of gas initially confined within one ring and at a specific azimuth, will be spread in the next snapshot in a way that follows the rotational motion of the disc. That is to say, the parcel is stretched in the azimuthal direction. Together with radial motion ascribed to bulk flows and/or diffusion, this creates an arc like feature in planar configuration space. This is illustrated in Fig.~\ref{mvFig}, which shows how tracers spread out in the x-y plane from an initial ring, centered at 8 kpc from the galactic centre, over the subsequent three snapshots. We can see from the figure that, after three snapshots, there is considerable radial movement of the tracers spreading both inwards and outwards from the initial ring boundaries.

We can quantify this effect of gas redistribution by constructing the histogram of the new radial position of tracers at snapshot $n+1$. Initially, at snapshot $n$, the distribution of tracers is approximately a top hat function with the width of the ring and median at the centre of the ring. At the next snapshot, the movement of tracers outside the ring leads to a new distribution with an different width (usually larger) and a shift (inwards or outwards) of the median of the distribution. We can directly utilize the information of the distribution at $n+1$ to describe the radial motion of the gas, using the difference between the new and initial median as a measure of the bulk radial motion and the width of the distribution as the measure of the spread of values around the new median. The caveat with this approach is the introduction of a floor in the value of the width because the width of the top hat distribution at snapshot $n$ is inherently included in the width of the distribution at snapshot $n+1$. This can become more problematic at the inner rings where gas is naturally more constrained in its radial motion. To avoid the presence of a floor value we can alternatively look at the distribution of tracers expressed by the difference in their initial and final galactocentric radii, by computing $\Delta R = R_{final} - R_{initial}$ for each individual tracer. Then by construction the initial distribution at snapshot $n$ is a delta function at $\Delta R = 0$ and the distribution at $n+1$ is a histogram centered at the new median with its the width similarly measuring the spread around the median, unconstrained from of a floor value. By testing both approaches we find that the resulting values for the widths are comparable, apart from the innermost rings, so the effect of the width of the ring does not appear very pronounced in the spread of the tracers between the two snapshots. Nevertheless, it is more reliable to use the histograms of $\Delta R$ in our analysis, eliminating the possible effect of the width of the rings on the results. In Fig. \ref{tracerhists} we show the histograms both in terms the galactocentric radii of the tracers and the difference $\Delta R$.

\subsubsection{Tracking gas motions}\label{sec:Tracking gas motions}
In our analysis, we exclude the tracers that in the time between the two snapshots have been in the wind phase. Although wind particles are launched in random directions in the Auriga wind implementation, the enhanced matter density in the plane of the disc restricts the outflows mainly in the perpendicular direction to the disc in a fountain flow. As a result, between the two snapshots a tracer can be launched from an inner ring in fountain trajectory and re-deposited in an outer ring. Hence, tracers that have been or are in winds may contaminate the information about pure radial motions within the plane. Tracers that have entered a wind particle are removed only for the snapshot pair but once they have returned to the disc later they may be included again as long as they have not entered a wind particle between the next pair of snapshots.

We further clean the sample by removing the data for rings belonging to halos that are in a merger state or more generally experiencing interaction with a satellite subhalo at a given snapshot. We choose 1/50 as the limit for the subhalo-to-central total mass ratio for a merger of importance. Merger cases are excluded on the reasoning that the disruption of the disc in the merger process can be significant enough that the cylindrical symmetry is lost and assigning rings cannot accurately represent the geometry of the gas motions. Mergers with the central galaxy can be identified using the SUBFIND \citep{Springel01} catalogs that are available for the simulations. We remove the snapshots at which the merger occurs according to SUBFIND and 3 snapshots, or equivalently $\sim$ 180 Myr, before and after the merger to partially account for the tidal interactions in the gas and stellar distribution that happen during the merger process and the time for the disc to settle after the merger. Changing the merger ratio limit to higher values (eg 1/10) does not influence significantly the results owing to the fact that most of the halos in the last 6 Gyr have very quiet merger histories and there are not many mergers in the mass ratio range 1/10 to 1/50. The number of halos we study is small enough that we have also visually checked the positional distribution of the gas tracers between the different snapshots and confirmed that this method successfully removes periods of significant disturbance by mergers. Filtering out the snapshots during merger phases removes 30 per cent of the total rings in the sample.

In Fig. \ref{tracerhists}, we choose to demonstrate characteristic examples of histograms obtained for two rings in the same halo, one inner and one outer one. We find that the histograms tend to be reasonably symmetric around the new median position of the gas, \ie{}gas tracers travel both inwards and outwards in the radial direction by roughly the same amount. In the majority of cases, the histograms can be accurately fit by a Gaussian function and we can use the standard deviation of the Gaussian to approximate the width of the distribution. However, there are cases for which the distribution of tracers in the next snapshot is not well approximated by a Gaussian (see example in Appendix Fig. \ref{badhist}). These cases arise almost exclusively in the outer rings of discs, which are more susceptible to external interactions (from subhalos) or mixing with the newly accreted gas because of their lower surface density. Furthermore, in the case of mergers, we observe more irregular distributions because the incoming subhalo can disturb the outer regions of the disc, leading to histograms that appear skewed or more random with large amounts of material having moved much further inwards or outwards. Skewed distributions are mostly eliminated by the merger cut.

Due to the possibility of such asymmetric distributions, we prefer to use the percentile ranges in order to describe the width of the distribution in this work. The 16-84 percentile range in particular is useful for evaluating the goodness of Gaussian fitting. If the histogram resembles a Gaussian, then the 16-84 percentile range should be very similar to twice the width of a Gaussian fit, $2\sigma$. We find that in most cases the two quantities can be used interchangeably, as shown in Fig. \ref{perc_gauss} in the Appendix. 

We thus extract two quantities from the shape of the histograms: the 16th-84th percentile divided by 2, which we will refer to as the `width', $w$, in kpc; and the difference in median galactocentric radius between the initial (at snapshot $n$) and final (at snapshot $n+1$) tracer distribution, which we will refer to as the `median shift', $\Delta \mu$, in km s$^{-1}$ (\ie{}normalising by the time difference $\Delta t$ between the snapshots).

There is a potential caveat that, to perform this kind of analysis, we ideally need to have a large number of tracers in a given annulus. Annuli with an insufficient number of tracers can contaminate the sample by mere lack of statistics, which leads to low confidence in the measurement of the percentile range. This becomes a problem usually in the outermost rings, where the density of cold/star-forming gas is low. Therefore, in this work we only consider annuli with a minimum of 500 tracers at snapshot $n$. This cut only removes 0.8 per cent of the rings.

We repeat the above process between all pairs of consecutive snapshots. This gives us a set of data for each ring that is its radius, its initial snapshot, the spread, and the median shift,
\begin{equation}
\tn{Ring}\,(\tn{halo}_{j},\,r_{i},\,t_{k},\,w_{ijk},\,\Delta\mu_{ijk})\ \ 
\end{equation}
where $\tn{halo}_{j}$ is the $j$th halo to which the $i$th ring belongs at the $k$th snapshot. Carrying out the analysis for the 14 halos, splitting each disc into 20 rings and working over 100 snapshot pairs, provides 28000 data points in the raw sample. We use the 100 last snapshots of the simulation , which is a total of lookback time of approximately 6 Gyr.

Furthermore, each ring has a set of associated properties that can be measured, such as the gas surface density $\Sigma_{\rm gas}$, total surface density $\Sigma_{\rm tot}$, gas fraction $f_{\rm gas}$, velocity dispersion $\sigma_{\rm tot}$ (as well as in individual directions $\sigma_{\rm r}$, $\sigma_{\rm z}$), the Toomre Q parameter for the gas $Q=(\sigma_{gas}\kappa_{gas})/(\pi G \Sigma_{gas})$, $\kappa$ being the epicyclic frequency and $\sigma_{gas}$ the total gas velocity dispersion using all three spatial components, and finally the star formation rate. These quantities can be extracted from the tracer particle data which inherit their properties from their parent gas cells. The velocity dispersion is calculated using the individual velocities of each tracer in the gas phases. The surface densities, are computed by counting the number of tracers in the gas phases ($\Sigma_{\rm gas}$) and stars and gas phases ($\Sigma_{\rm tot}$), multiplying by the associated masses and dividing by the surface area of the ring. In addition, we calculate the accretion rate onto a given ring $\dot{M}_{\rm acc}$ and the accreted mass fraction ,that is the accreted mass divided by the gas mass already present in the ring, $f_{\rm acc}=M_{\rm acc}/M_{\rm gas}$. The accreted mass is calculated by counting the tracers which are in the gas phases (non-SF and SF) and which at snapshot $n$ are outside the ring limits and at snapshot $n+1$ within them. This is strictly accretion of material that is external to the defined disc region at the initial snapshot and does not include material exchange between different rings. The accretion rate is then given by the total mass of accreted tracers divided by the time between the two snapshots, $\dot{M}_{\rm acc}=M_{\rm acc}/\Delta t$. We also divide the accreted mass fraction by the snapshot spacing to get a time-normalised quantity: $\dot{f}_{\rm acc}=f_{\rm acc}/\Delta t=\dot{M}_{\rm acc}/M_{\rm gas}$. The quantity $\dot{f}_{\rm acc}$ is essentially the inverse of an accretion timescale.

\subsubsection{Evolution over time}\label{sec:Evolution over time}
\begin{figure}
\centering
\includegraphics[width=\linewidth]{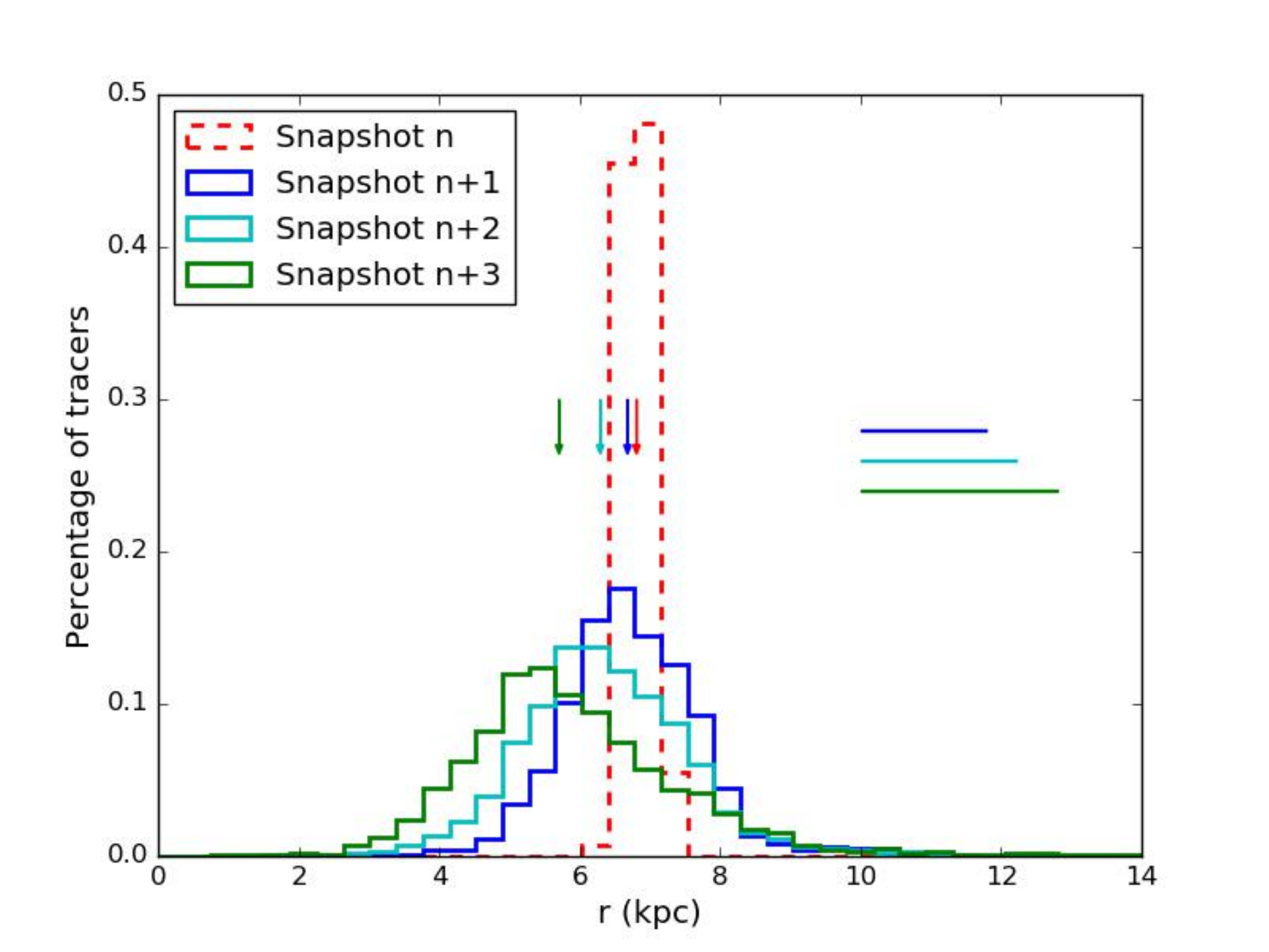}
\caption{Radial distribution of the tracers selected in a ring initially at snapshot n as it evolves at subsequent snapshots $n+1$, $n+2$ and $n+3$. The arrows show the medians of the distributions and the horizontal lines the 16-84th percentile.}
\label{timeHists}
\end{figure}

In the fiducial case, we calculate $w$ and $\Delta \mu$ between consecutive snapshots (\ie{} between snapshots $n$ and $n+1$), but we can equally compute them for the time between snapshots \textit{n} and $n+2$ or $n+3$. In these cases, the time difference is roughly two and three times longer, so the histograms appear naturally broader. The quantity $w$, as expressed in kpc, is therefore dependent on different timestep or snapshot spacing selections. By looking at the evolution of $w$ in a given ring between $n+1$, $n+2$ and $n+3$, we can identify its time dependence, assuming it follows a proportionality of $w \sim \Delta t^{a}$, where $\Delta t$ is the time difference between the two snapshots. This is important if we want to have our parametrised quantities in a timestep invariant form, so that the result can be generally applied to models or simulations with different timestep widths. In Figure \ref{timeHists}, we show an example of how the radial positions of a group of tracers in a given ring have evolved after 1, 2 and 3 snapshots. We stop at 3 snapshots after snapshot n, which is a time interval comparable to the dynamical time of the disc for most radii, because is sufficient to capture the radial flows that we want to study. Using n+4 or n+5 gives convergent results in the radial and time evolution of w and $\Delta \mu$. If we proceed further, the histograms deviate from a Gaussian distribution, losing a clear peak. In addition as we use larger time difference we increase significantly the error on the measurement of the quantity $w$.

In section \ref{sec:Results}, we provide the exact time dependence of $w$ and how different snapshot spacings influence it and $\Delta \mu$.

\subsubsection{Redshift and mass dependencies}\label{sec:Redshift and mass dependencies}
In order to check if there is any significant redshift dependence to the radial flows studied here, we have initially split all the output snapshots into three broad time bins of 2 Gyr. Each bin contains approximately 30 snapshots, for which we calculate the tracer positions at all the snapshot pairs $n$ and $n+1$. We find that the there is no significant redshift evolution in the trends that we present in Section \ref{sec:Results}. Furthermore, we have split the sample between the seven most massive and least massive halos, but find no evidence for any mass dependence. Therefore, for our final study we combine the data over the last $\sim$6 Gyr (100 snapshots) for all the halos.

\begin{figure*}
\centering
\begin{minipage}{.49\textwidth}
  \centering
  \includegraphics[width=\linewidth]{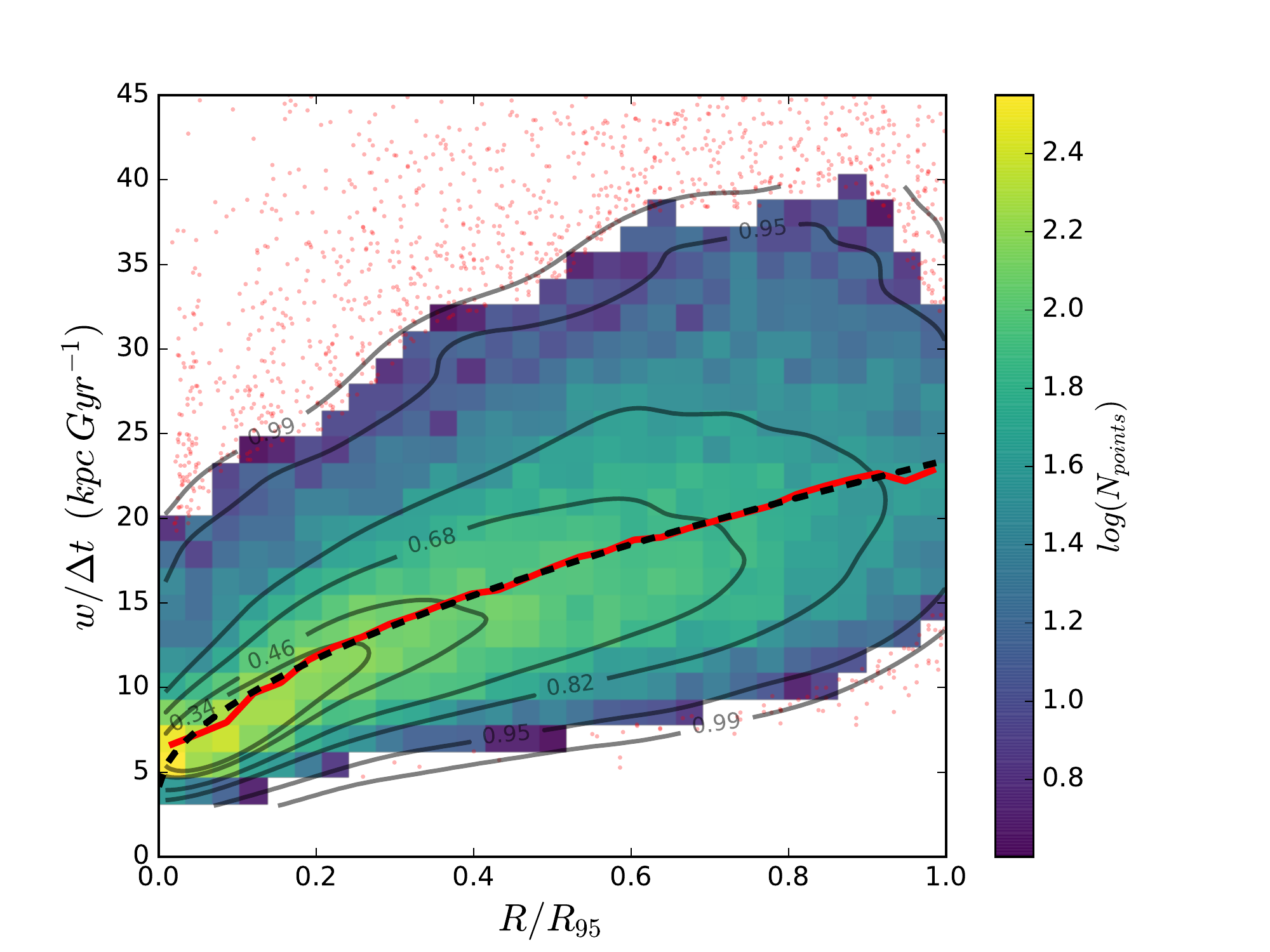}
\end{minipage}
\begin{minipage}{.49\textwidth}
  \centering
  \includegraphics[width=\linewidth]{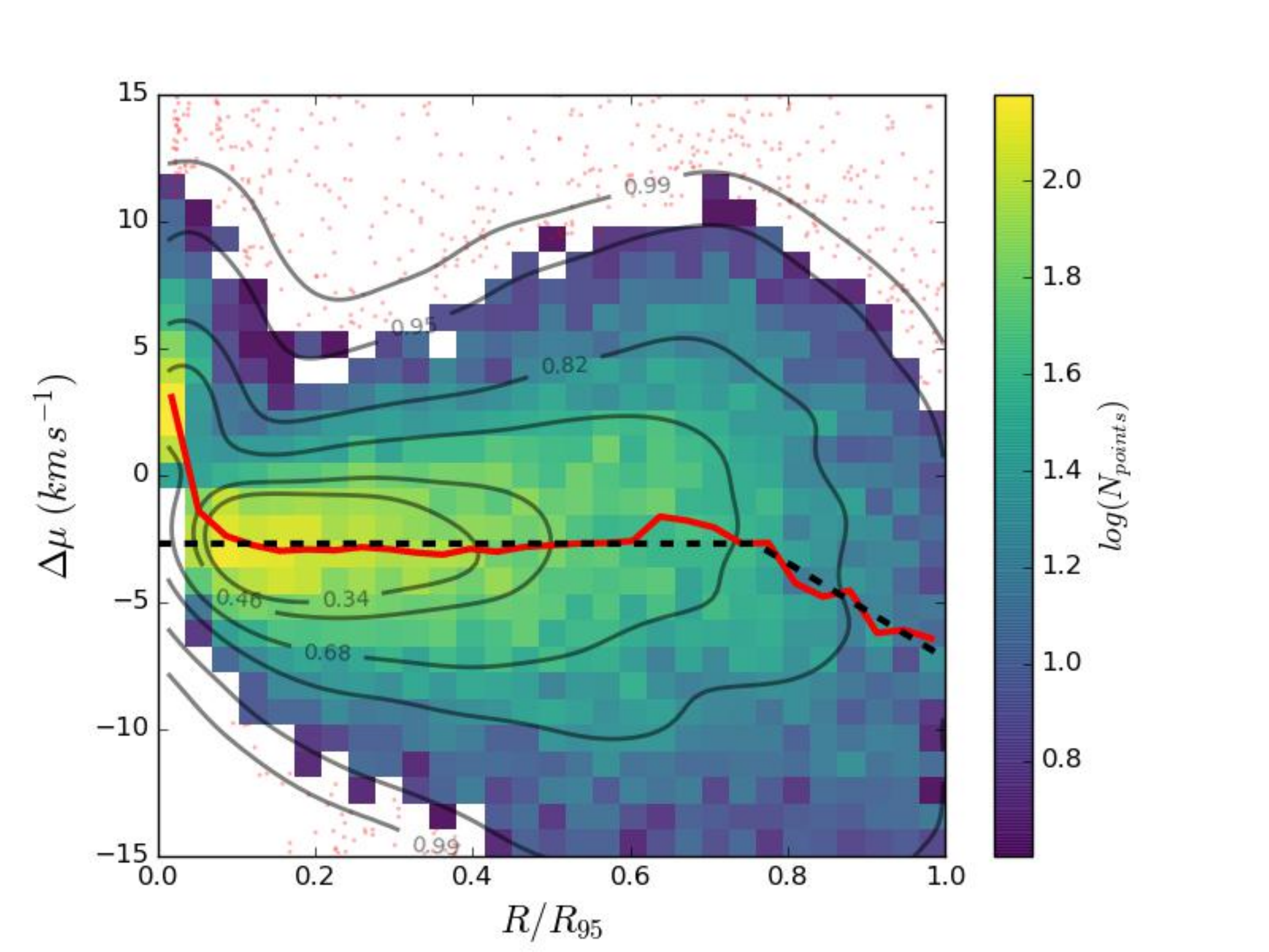}
\end{minipage}
\caption{Radial dependence of the spread $w$ (left) and the median shift $\Delta \mu$ (right). The data points, for all 14 halos over the range of 6 Gyr, are represented as a number histogram and the red line shows the median curve. Also shown the contours enclosing a given percentage of the points. For these plots, we evaluate the quantities using consecutive snapshots with time difference 60 Myr on average. We observe an increasing trend with the radius for $w$ which can be fit with a power law. In this plot we present $w$ normalised by the snapshot spacing $\Delta t$ to account for the small differences in the exact spacing between the different snapshot pairs. For $\Delta \mu$ there is a flat trend up to 0.75 disc radii and a linear drop in the outer regions where there is faster inflow of material.}
\label{radmed}
\end{figure*}

\section{Results}\label{sec:Results}

The first observation that we naturally want to test is how $w$ and $\Delta \mu$ vary with the radial position of the ring. We find that $w$ is larger on average for rings at larger galactocentric radius. The median of the $r$ -- $w$ relation for the whole sample can be best fit with a power law of with slope $<1$, as shown in the left panel of Fig. \ref{radmed}. In the inner regions of discs, $\Delta \mu$ is a constant value of around $-3\,\rm{km\ s}^{-1}$ up to almost 70\% of the disc radius, in agreement with the observations that show gradual inflows of gas in disc galaxies (eg \cite{Schmidt16}). In the outer regions, the value of $\Delta \mu$ becomes more negative, ranging between -3 and -15 $\rm km \, s^{-1}$ on average, indicative of enhanced gas inflow. In the very centre of galaxies, positive (outwards) values of low speed are a manifestation of the fact that the gas in the innermost ring cannot travel any further inwards but also that higher outflow speeds are driven by central AGN feedback. 

The above statements are visualised in Figure \ref{radmed}, which displays the compilation of data for all the halos over the selected rings (excluding merger cases and low number of tracers, as discussed in Section \ref{sec:Tracking gas motions}) and over the aforementioned snapshot range.

These statements hold true if we average the data for all halos (as shown in Fig. \ref{radmed}) but also if we look at each halo individually. For an individual halo, the curve of $w$ and $\Delta \mu$ versus radius can be less smooth in some cases, although the radial trends are still similar. We find that before removing the merging stages, halos with quieter merger histories and a more stable disc evolution return more consistent results between different time intervals.

For three of the halos from the higher mass sample ($1-2 \times 10^{12} M_{\odot}$) in our simulation suite, we measure high $w$ and irregular $\Delta \mu$ values at inner radii. Looking directly at the cold gas tracer x-y plane for these halos, we see large holes devoid of gas in the inner regions which have bubble like profiles. These holes are created by feedback from the AGN, which pushes gas out of the central region, increasing the $w$ measured and giving positive (outwards) $\Delta \mu$ values in these rings. We mitigate these feedback effects by removing the tracers that have been in wind particles, but the overall feedback effect cannot be removed completely. However, these bubbles are only present in a small subset of the snapshots, so do not influence our conclusions statistically.

We test for the convergence of the results by varying the number of radial bins and the height cut. In the first case, if we use a very small number of rings (\eg{}5-7, compared to the 20 rings we use by default), we get higher values for the spread at a given radius. Using more than 25 is oversampling and results in a low number of tracers per ring. In general, we get convergent results if between $\sim{}10$ and 25 rings are used. Varying the maximum height above and below the disc plane between 2 and 4 kpc does not have any qualitative effect on the median trend, although there is no convergence if we use a very conservative height cut ($<$1 kpc), because not all the tracers that are relevant for disc flows are included. 

\subsection{Timestep invariant expression of \texorpdfstring{$w$}{w} and \texorpdfstring{$\Delta{}\mu$}{Delta mu}}\label{sec:power_evol}

\begin{figure*}
\centering
\begin{minipage}{.4\textwidth}
  \centering
  \includegraphics[width=\linewidth]{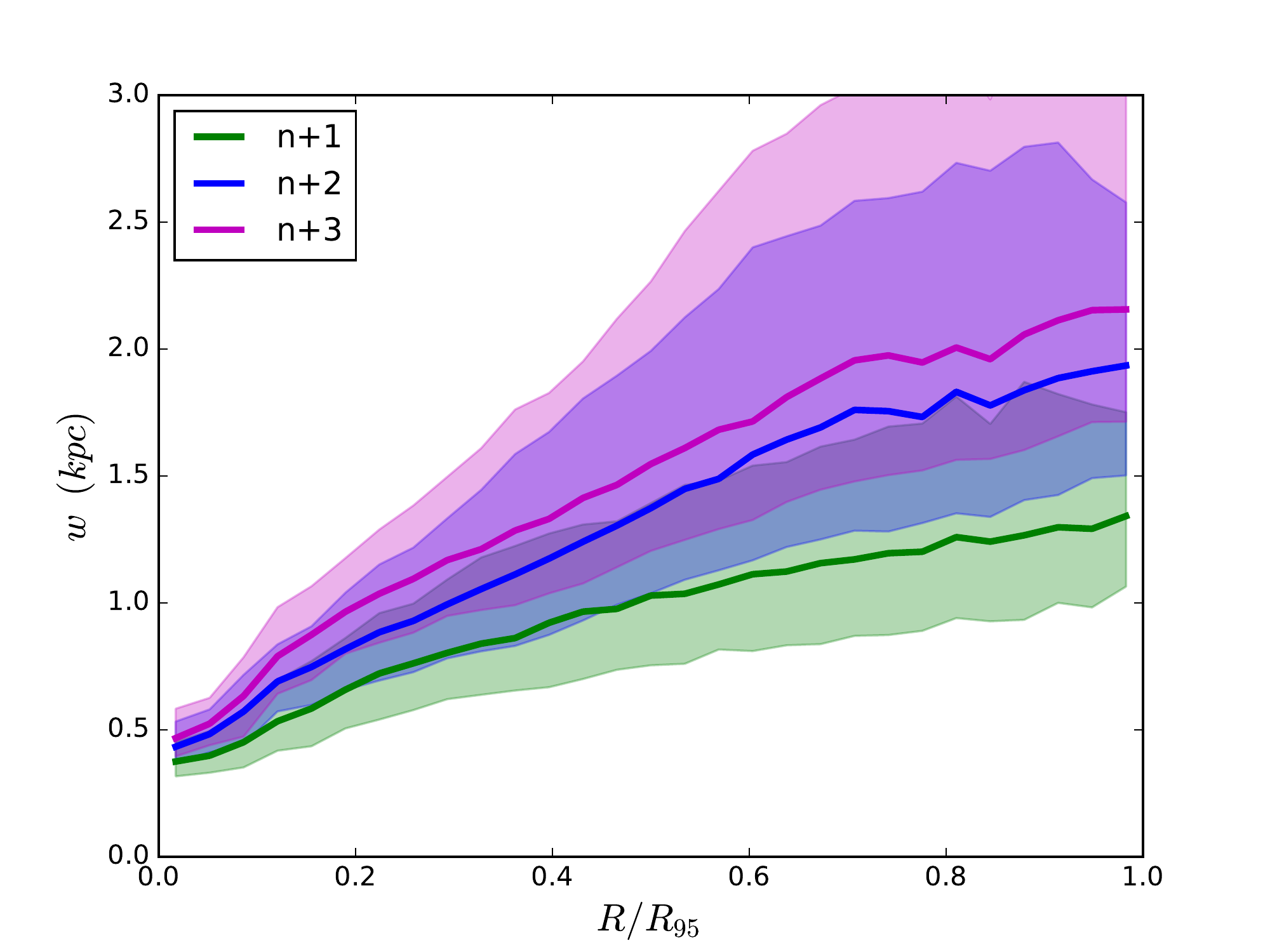}
\end{minipage}
\begin{minipage}{.4\textwidth}
  \centering
  \includegraphics[width=\linewidth]{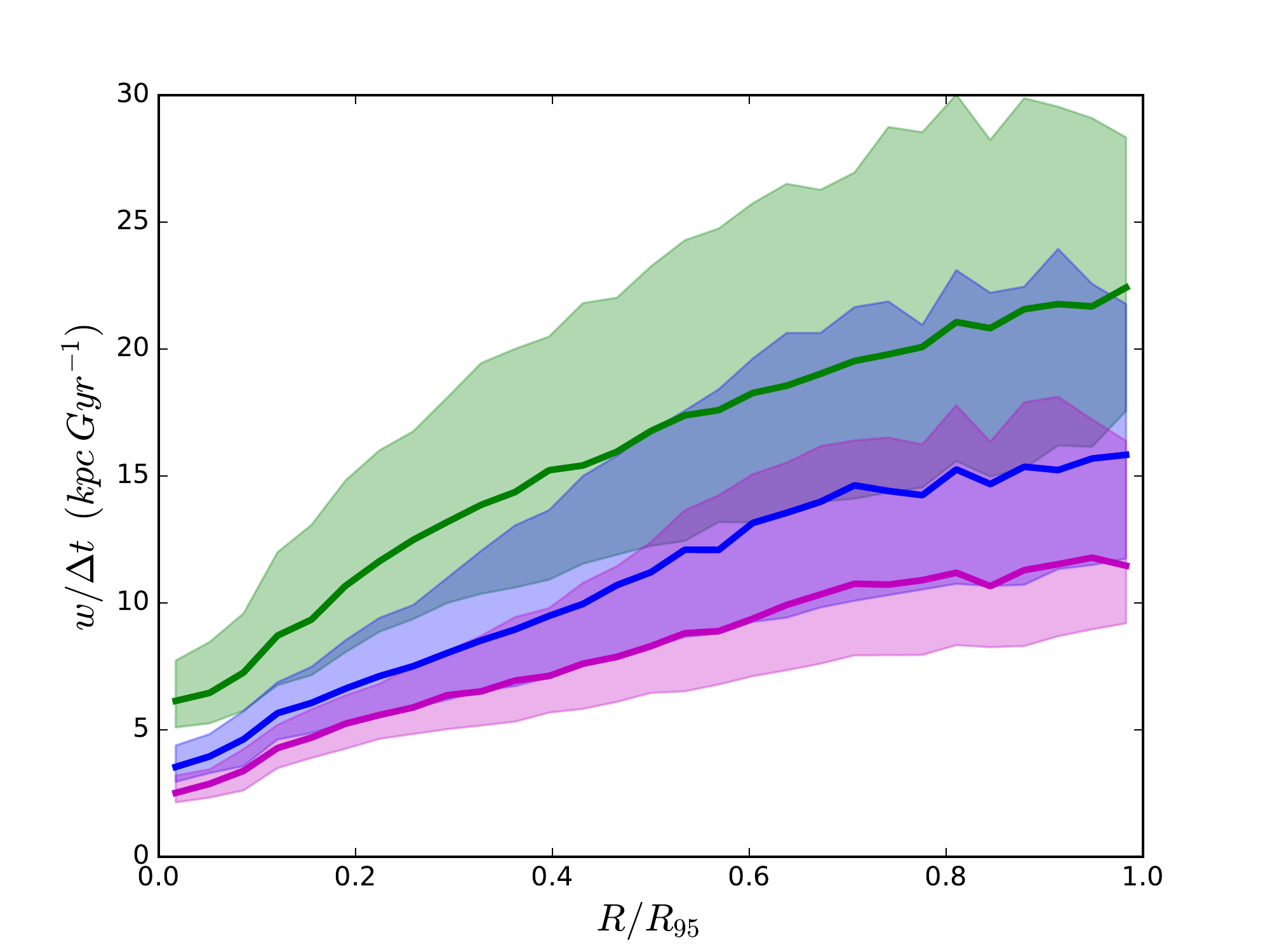}
\end{minipage}
\begin{minipage}{.4\textwidth}
  \centering
  \includegraphics[width=\linewidth]{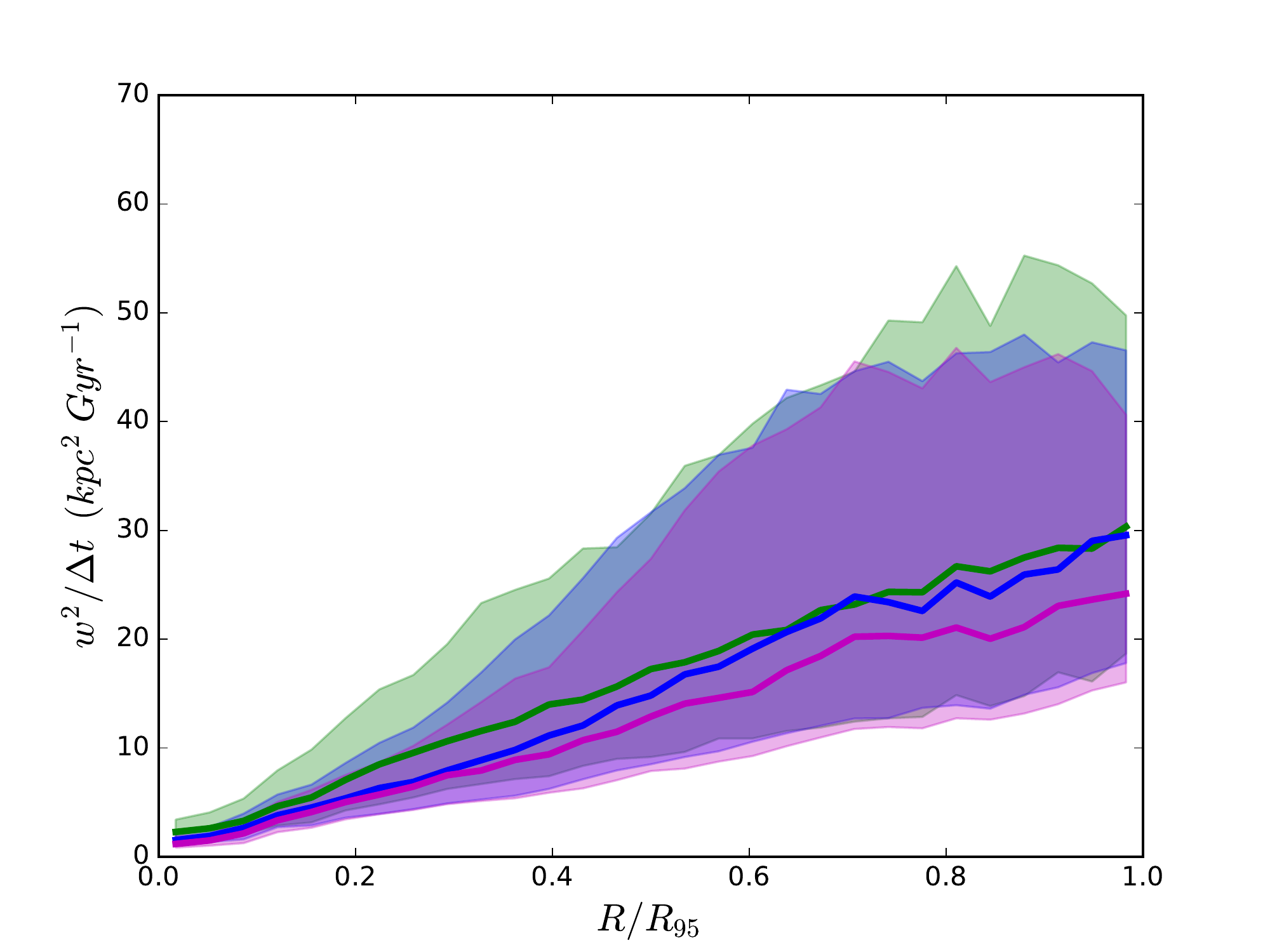}
\end{minipage}
\begin{minipage}{.4\textwidth}
  \centering
  \includegraphics[width=\linewidth]{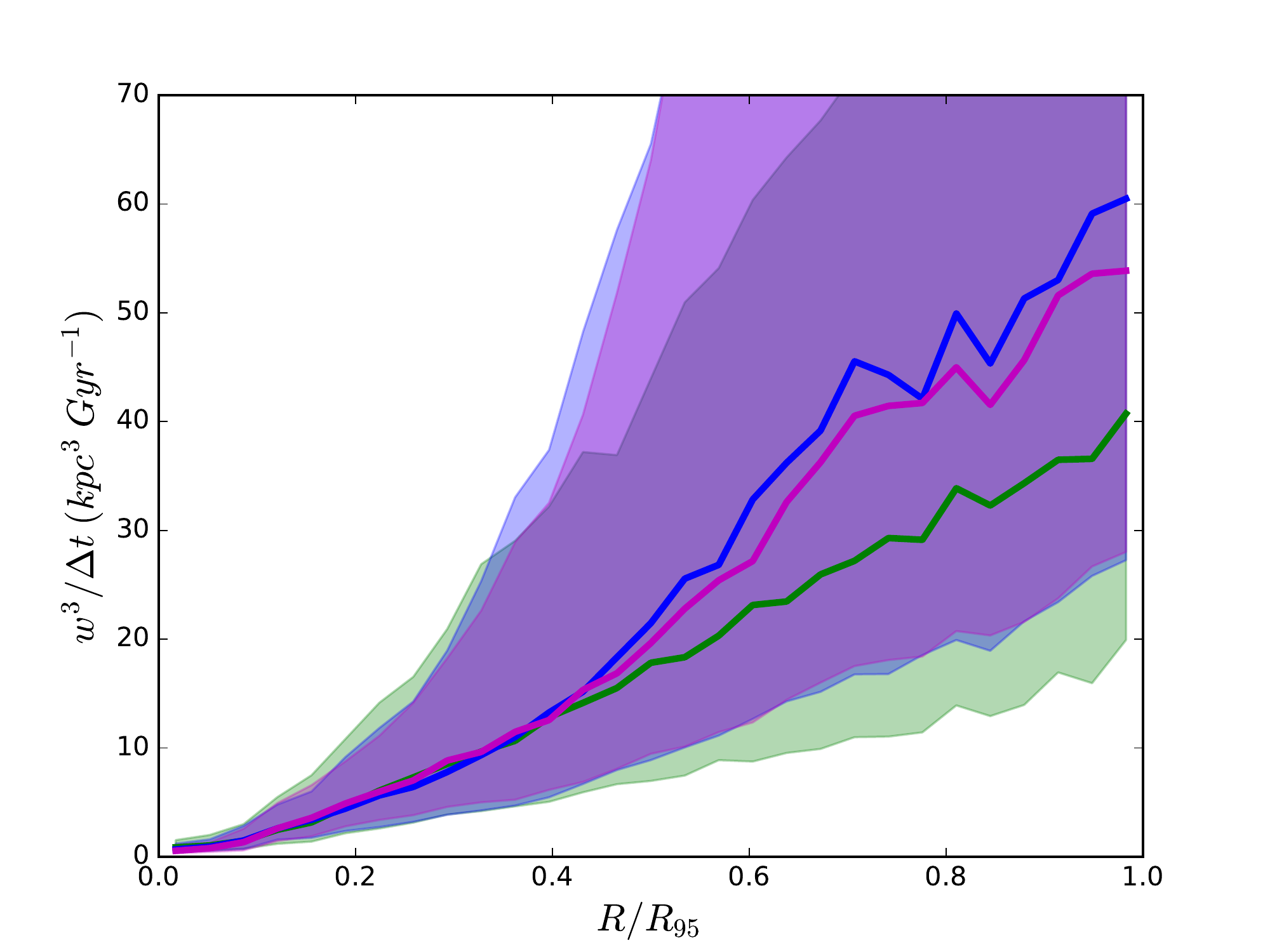}
\end{minipage}
\begin{minipage}{.4\textwidth}
  \centering
  \includegraphics[width=\linewidth]{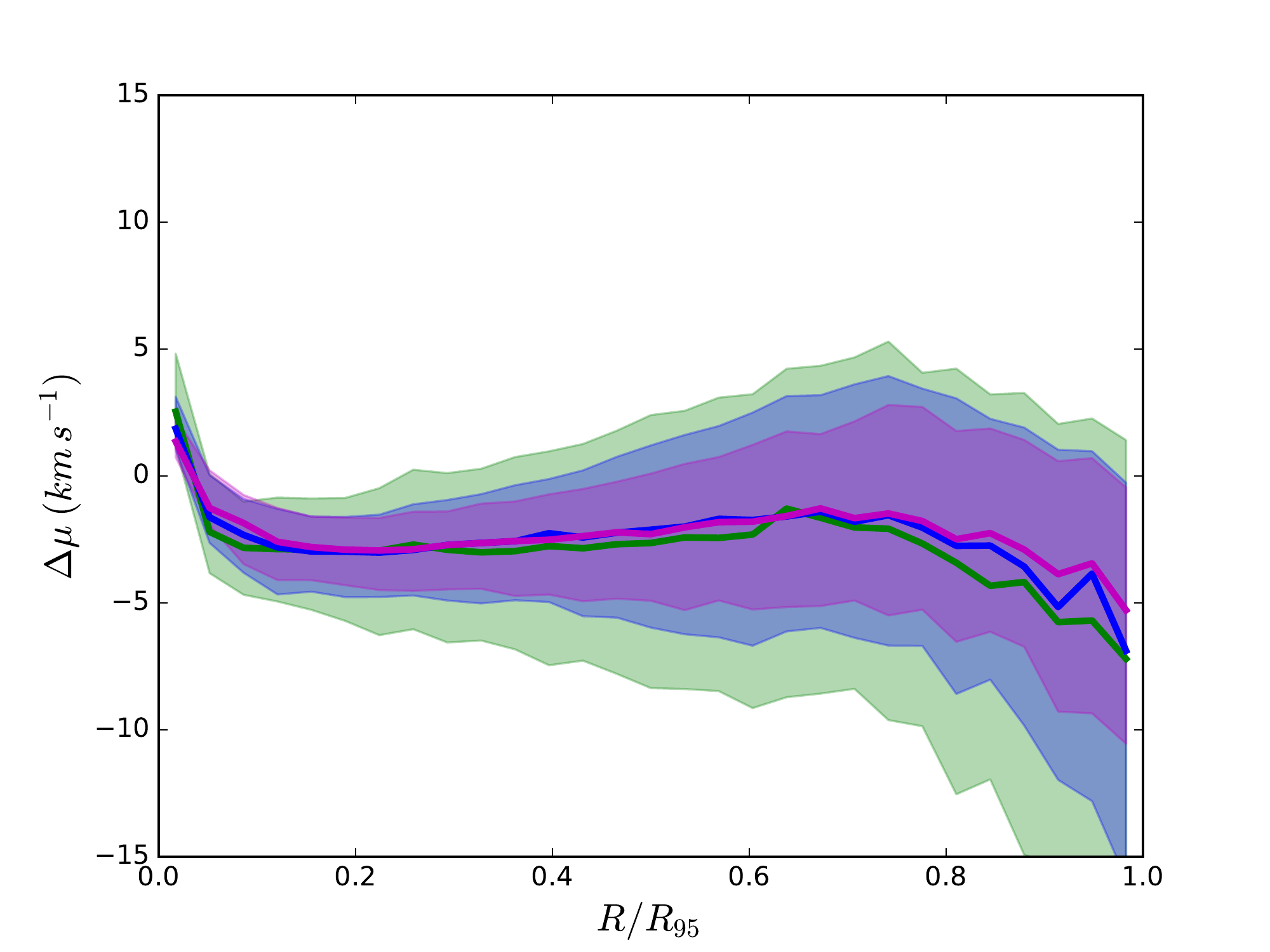}
\end{minipage}
\caption{Mean radial profile of $w$ (top left), $w/\Delta t$ (top right), $w^{2}/\Delta t$ (middle left), $w^{3}/\Delta t$ (middle right) when calculated between snapshots $n$ and $n+1$ (green), $n+2$ (blue) and $n+3$ (magenta). $w$ naturally has higher values for the larger timestep case, between $n$ and $n+3$. The quantity $w^3/\Delta t$ is the one that leads to the best convergence between the three cases in contrast to a simple $w/\Delta t$ expression which does not represent the time evolution of w accurately or $w^2/\Delta t$ which shows a systematic dependence. Bottom right panel: The radial profile of the median shift $\Delta \mu$ is similar in all three cases showing a consistent calculation of the bulk inflow velocity. The shaded regions show the 1$\sigma$ intervals around the median curves.}
\label{w_timevol}
\end{figure*}

As mentioned above, in Fig. \ref{radmed} we present the quantities $w$ and $\Delta \mu$ as calculated between two consecutive snapshots in the simulation. The time difference, $\Delta t$, between consecutive snapshots is on average 60 Myr, with a range between $\sim$50-70 Myr. Given this, when looking between snapshot $n$ and $n+2$ or $n+3$, $\Delta t$ increases to $\sim{}100-140$ Myr and $\sim{}150-210$ Myr, respectively. $\Delta \mu$ is presented in units that already account for such differences in $\Delta t$, but this is not the case for $w$. In Fig. \ref{w_timevol}, we show how $w$ and $\Delta \mu$ vary on average if calculated between snapshots $n$ and either $n+1$, $n+2$, or $n+3$. For $\Delta \mu$, we find a convergence in the results around the value of $-3\,\rm{km\ s}^{-1}$. For $w$, a dependence on the number of snapshots chosen is clear. From the distance between the median curves in the top-left panel, we get an indication of how $w$ varies as we double or triple the size of $\Delta t$. The increase is not directly proportional to $\Delta t$, as shown in the top-right panel where we plot the quantity $w/\Delta t$. Instead, for the quantity $w^{2}/\Delta t$, corresponding to a $\sqrt{\Delta t}$ time dependence, we see better convergence within the scatter. However, we also find a systematic trend where $n+3$ lies lower on average than $n+2$, which in turn is lower than $n+1$. Whether the spread of the tracers was governed by a pure diffusion process, we would expect a time invariance with $w^2/\Delta t$. Finally, $w^3/\Delta t$ converges very well in all three cases in the inner disc, and the deviation in the outer parts shows no systematic (\ie{}the $n+3$ median line now lies in between the other two) so it is also consistent within the scatter. Hence, $w^{3}/\Delta t$ appears to be the quantity that is most timestep invariant when describing the spread of the tracers.

\begin{figure}
\centering
\includegraphics[width=\linewidth]{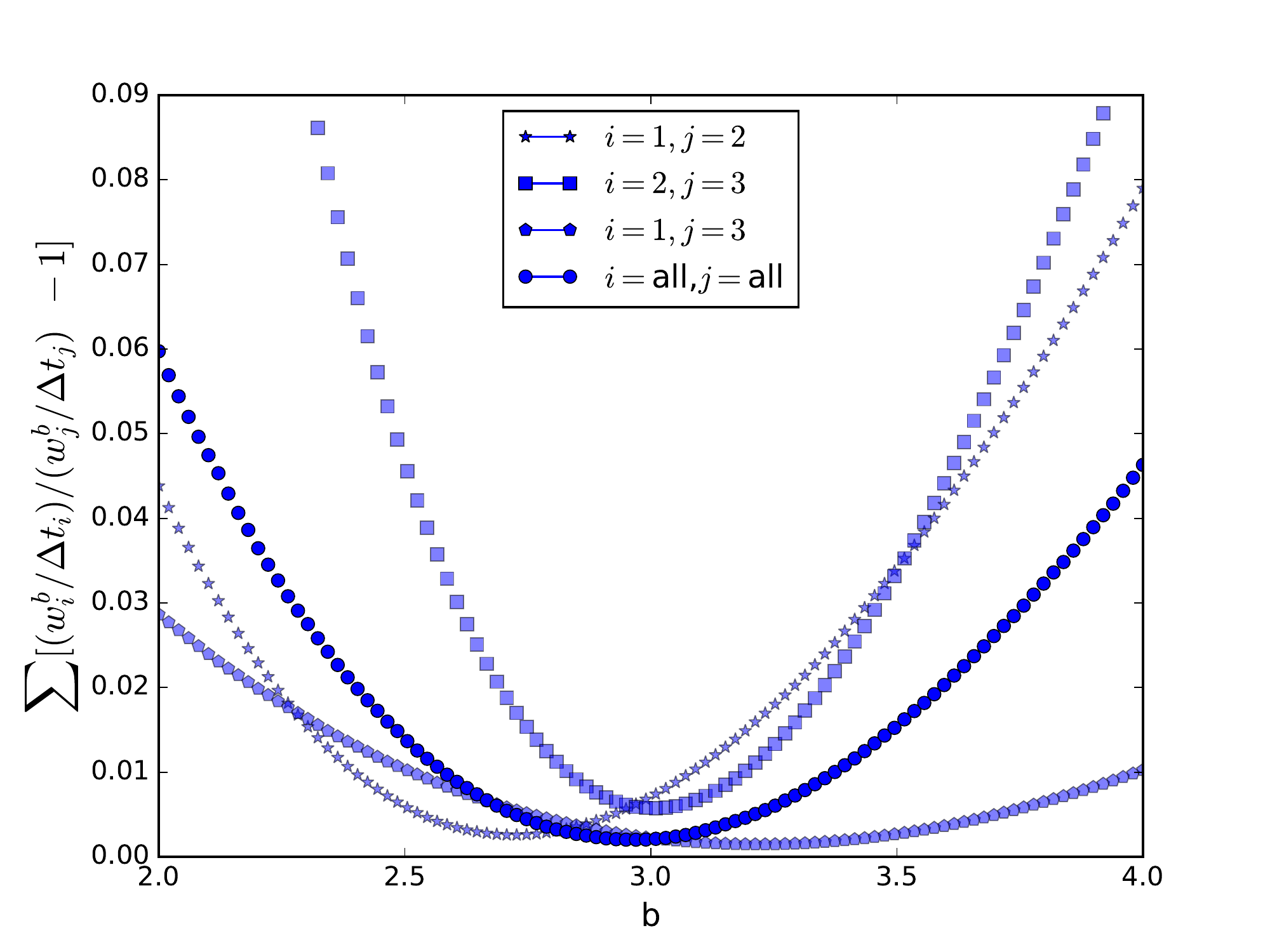}
\caption{The ratios described by Eqn. \ref{eqn:w_ratios} in Section \ref{sec:power_evol} help us identify the best fit value for the power in the expression $w^b/\Delta t$. The x-axis has a range of b values and the y-axis is the measure of the deviation of the ratios from 1 where a smaller value in the y-axis indicates a better fit around 1. The combined data (solid circular points) yield a minimum for the parameter $b$ at a value $b=3$, while also taking each ratio individually (semi-transparent points) gives us minima values around $b=3$.}
\label{minima_b}
\end{figure}

\begin{figure}
\centering
\includegraphics[width=\linewidth]{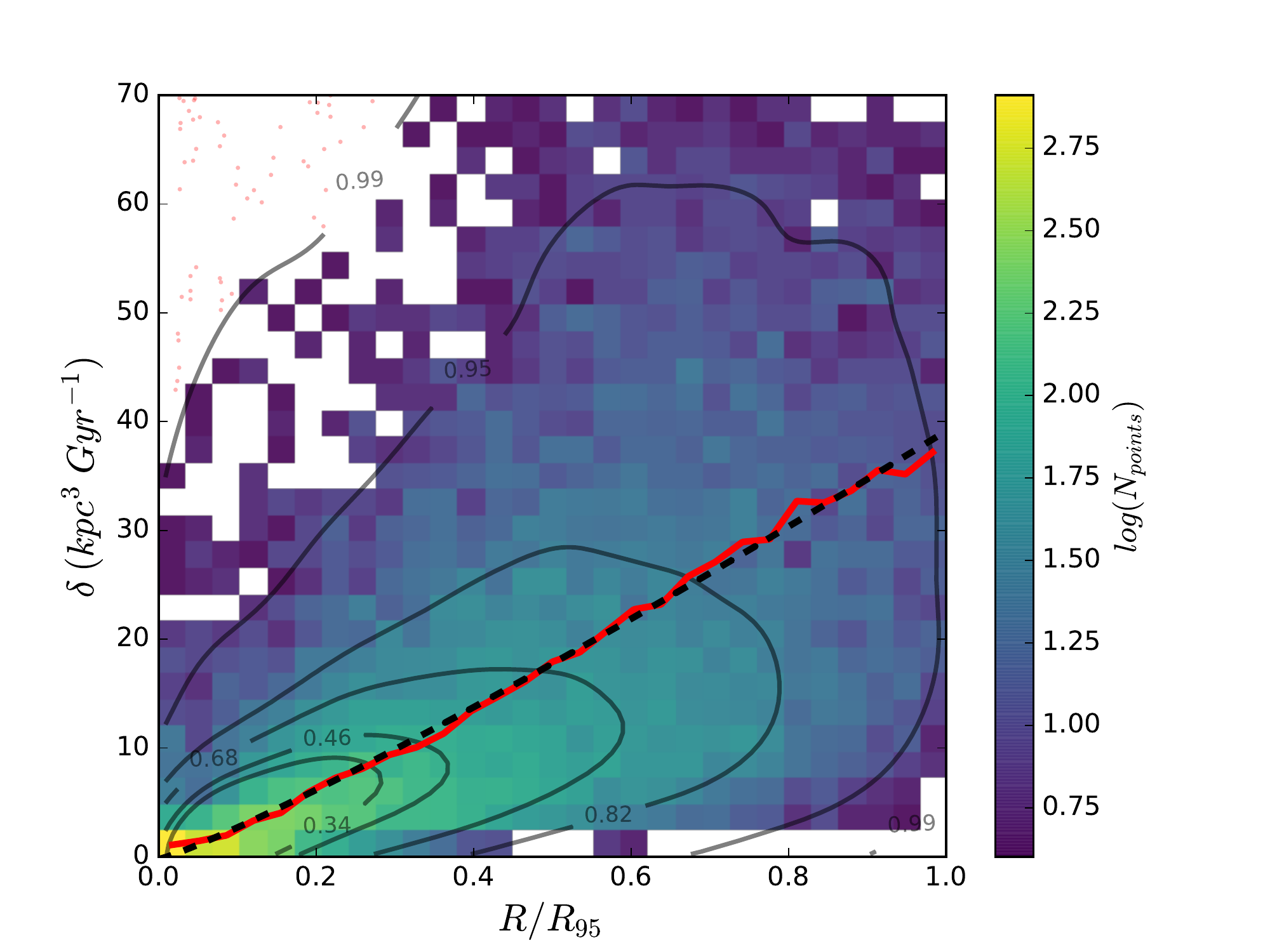}
\caption{Radial dependence of $\delta$, similar to Fig. \ref{radmed}, showing the best fit to the median relation. The best fit power is 1.1, slightly different to a linear relation in the inner radii.}
\label{delta_rdepend}
\end{figure}

We want to quantitatively confirm the cubic power dependence by running the following test. Based on the assumption that $w \sim \Delta t^{a}$, it follows that $w^{b} / \Delta t = \tn{const.}$, where $a=b^{-1}$, independently of whether $w$ is calculated between the pairs of snapshots $[n,n+1]$, $[n,n+2]$ or $[n,n+3]$. So, in order to identify the best value for the power $b$, which will show us how $w$ evolves with time, we calculate the following three ratios,
\begin{align}\label{eqn:w_ratios}
    \nonumber q_{12} = & (w_{1}^{b}/\Delta t_{1})\,/\,(w_{2}^{b}/\Delta t_{2}) \\
    q_{13} = & (w_{1}^{b}/\Delta t_{1})\,/\,(w_{3}^{b}/\Delta t_{3}) \\ \nonumber q_{23} = & (w_{2}^{b}/\Delta t_{2})\,/\,(w_{3}^{b}/\Delta t_{3})\;\;,
\end{align}
where the subscripts on the right-hand side 1, 2, 3 show, respectively, whether $w$ and $\Delta t$ have been calculated between $[n,n+1]$, $[n,n+2]$ or $[n,n+3]$ for the tracers in a given ring at snapshot $n$. We can also combine the data for the three ratios to include the information for all three timesteps that are examined. If $w^{b} / \Delta t = \tn{const.}$ holds, these ratios should ideally be equal to 1 for the value of $b$ that better describes the process of radial spreading. We thus identify the value of $b$ that minimizes the difference of
\begin{equation}
    \sum\sub{rings} (q-1)^{2}\;\;\;,
\end{equation}
where $q$ can either be each of the above ratios independently or the combined data for all three of them. 
The above sum is minimised very close to the value $b=3$ (exact value 2.97) when using all the data, as shown in Fig. \ref{minima_b}, indicating that the quantity $w^{3}/\Delta t$ is the most timestep invariant. When using the individual ratios the minimum values range around $b=3$ from 2.7 to 3.4. If we consider only the outer part of the disc ($r/R_{95}$>0.75) the minimum value for b is 2.7 or only for the inner part ($r/R_{95}$<0.75) $\tn{b}_{min}=3.2$. We will define $\delta = w^{3}/\Delta t$ for simplicity from now on. This will be the quantity we aim to paramterise along with $\Delta \mu$. In Fig. \ref{delta_rdepend} we show the radial dependence of $\delta$ for the whole sample of rings.

\subsection{Dependence on physical properties}

We would like to check whether the radial dependence of $w$ and $\Delta\mu$ are driven by some physical process, or are correlated with physical properties either of the individual rings or the galactic disc as a whole.

If any dependencies present are not due to a global disc property, \ie{}do not vary significantly among galaxies, then we can treat each ring as an independent data point no matter which halo it belongs to. Then, the premise is that the width of the histogram is driven by some local, internal property within the ring or process associated with it (for example the perturbing effect of a local feature such as a spiral arm). 
As mentioned before, we observe an increase in $w$ with increasing radius (see Fig \ref{radmed}). There is a considerable scatter in this relation, but the overall trend is clearer when taking mean values of the spreads for given radii.There is also large scatter in the relation of $\Delta \mu$ versus radius towards both negative and positive values, which become more pronounced in the larger radii.

The source of the scatter could be due to a lack of homogeneity among the halos or a dependence on a secondary parameter that could be either directly measurable in the simulation output or acting in between the snapshots. When separating the data between the different halos and reproducing the $w$ -- $r$ and $\Delta \mu$ -- $r$ relations for each, we find that their median relations lie very close to each other and hence we cannot attribute the scatter in the full dataset to halo variance.

We have chosen to examine a number of local properties that could potentially influence $w$. Firstly, we consider the total baryonic surface density ($\Sigma\sub{tot}$), the gas surface density ($\Sigma\sub{gas}$), and their ratio the gas fraction ($f\sub{gas}$). These properties can tell us whether there is a direct relation between the flows and the amount of material in the ring, as well as distinguish between the effect of gas and total baryonic mass. We also consider the gas velocity dispersion ($\sigma$), which is a measure of the internal kinetic energy of the material and of the amount of turbulence. This is further split into the velocity dispersion in the radial direction ($\sigma\sub{r}$) and that perpendicular to the disc plane ($\sigma\sub{z}$), in order to identify which is dominant. We also examine the effect of accretion, which has been postulated as a driver of radial flows, by computing the mass accretion rate ($\dot{M}\sub{acc}$) onto a ring and the accreted gas mass fraction ($\dot{f}\sub{acc}$). Finally, the star formation rate (SFR), which relates the energy deposition from stellar feedback to the gas that could drive flows and the Toomre parameter Q as a measure of the gravitational instability that, as mentioned before, has also been related to gas flows.

\begin{figure*}
\centering
\begin{minipage}{.49\textwidth}
  \centering
  \includegraphics[width=\linewidth]{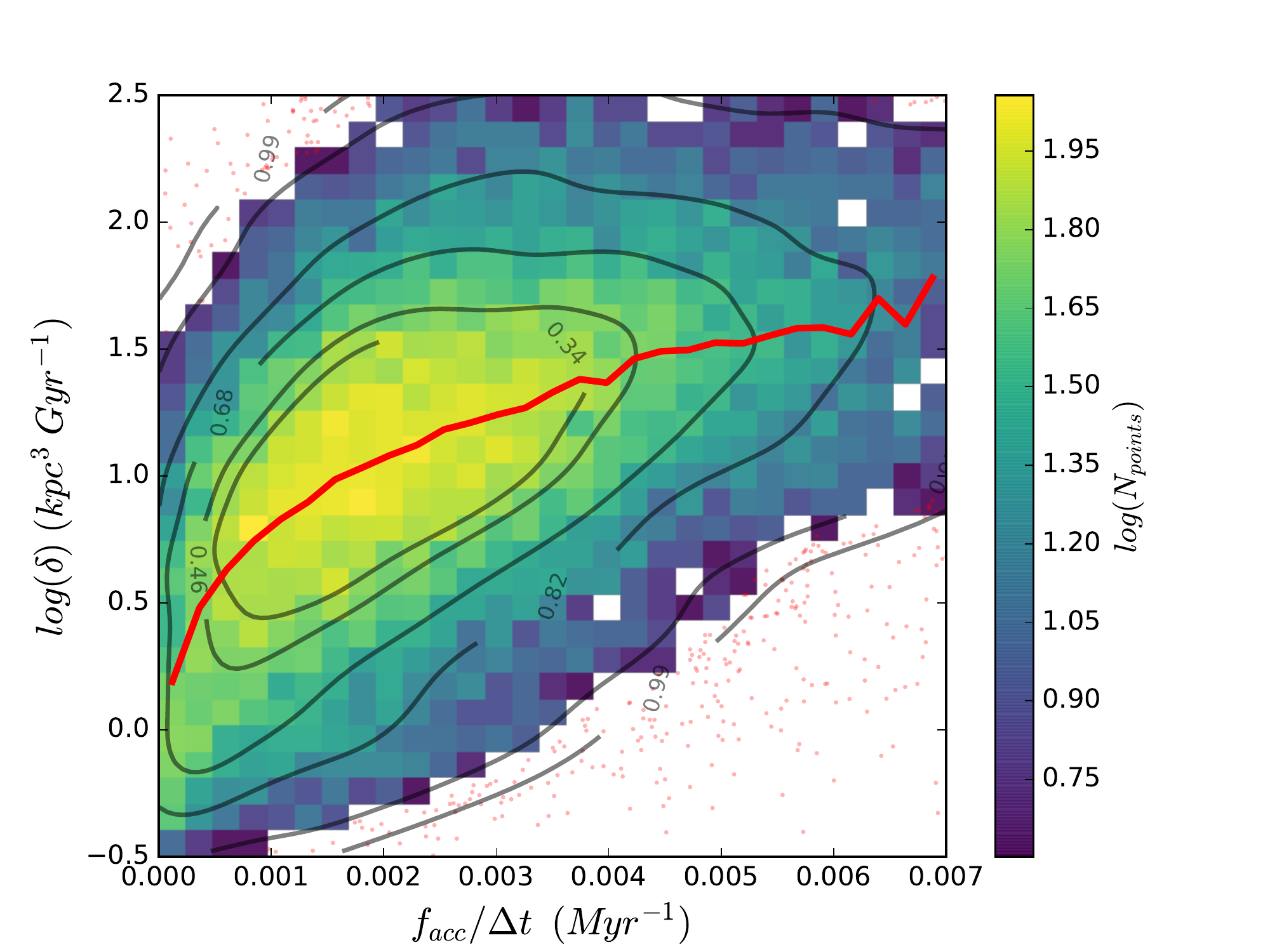}
\end{minipage}
\begin{minipage}{.49\textwidth}
  \centering
  \includegraphics[width=\linewidth]{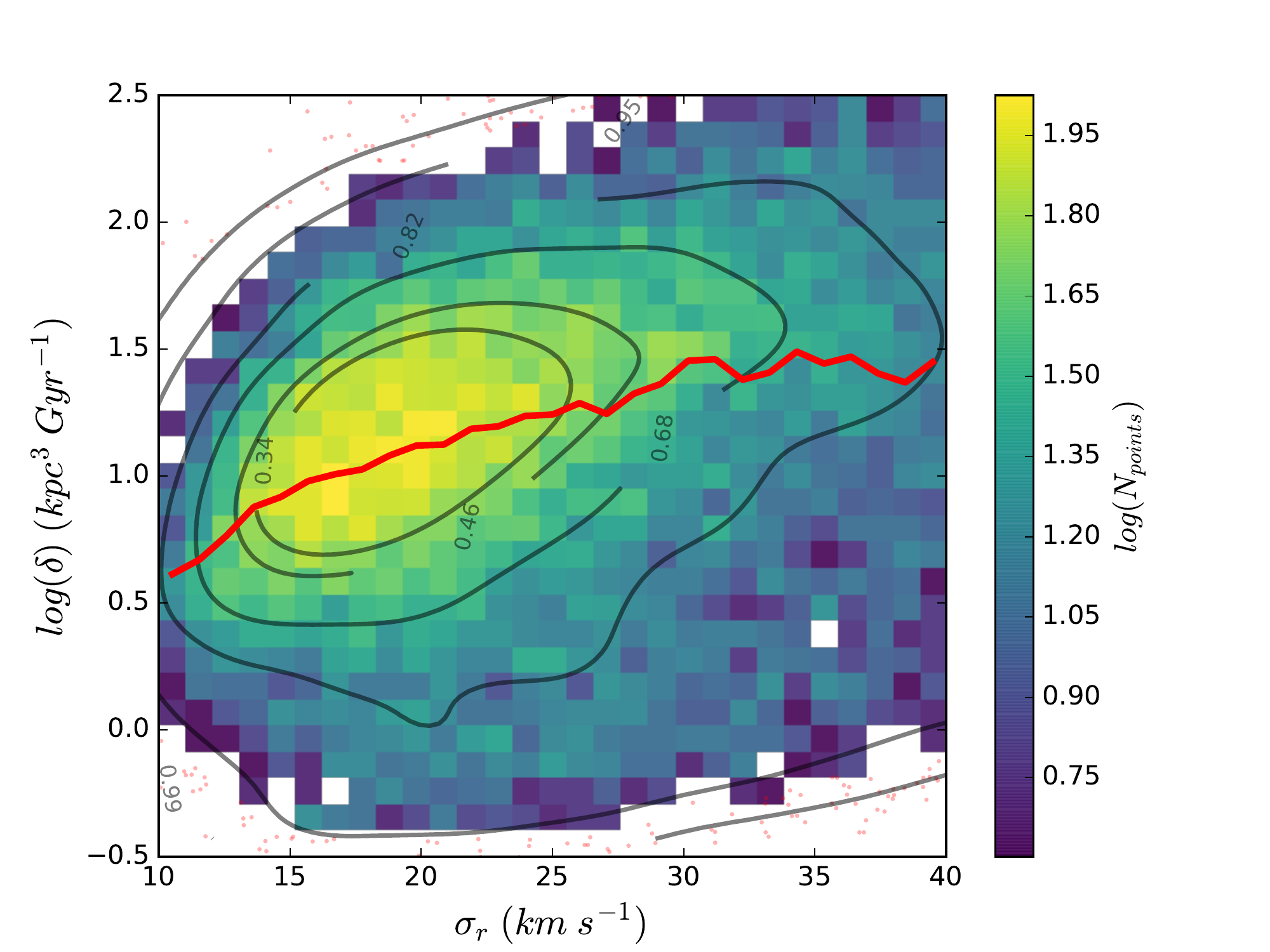}
\end{minipage}
\begin{minipage}{.5\textwidth}
  \centering
  \includegraphics[width=\linewidth]{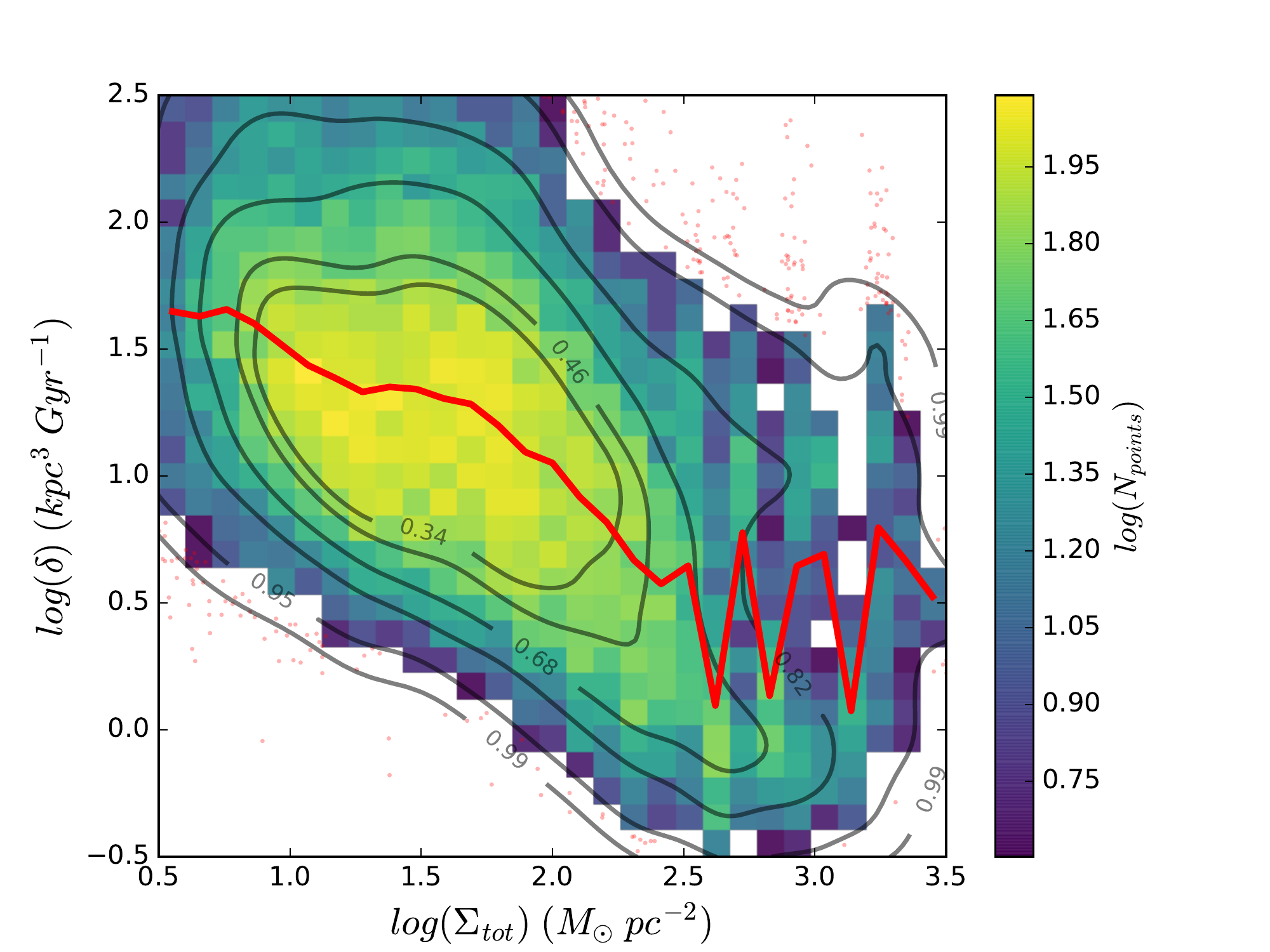}
\end{minipage}%
\begin{minipage}{.5\textwidth}
  \centering
  \includegraphics[width=\linewidth]{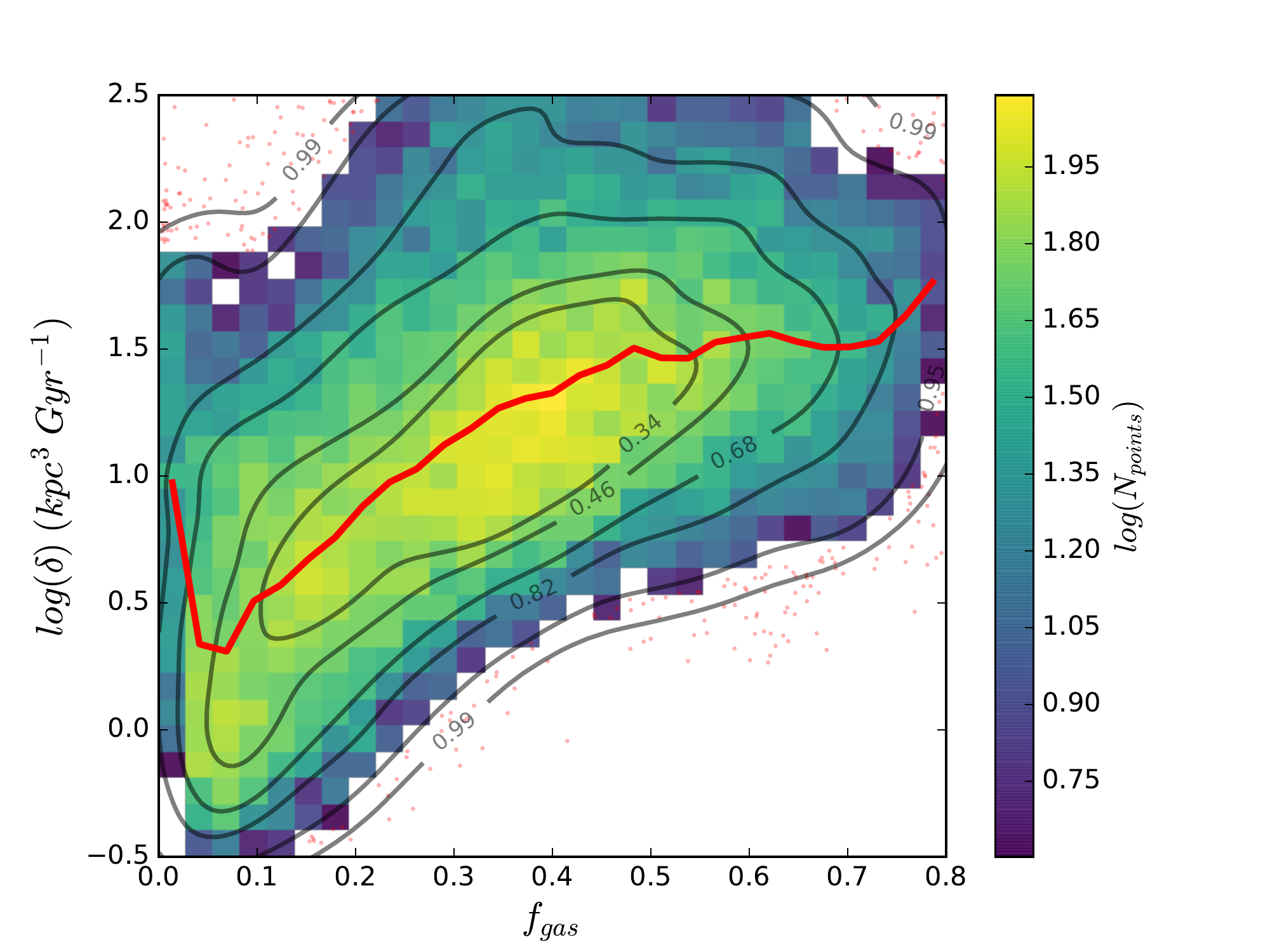}
\end{minipage}
\caption{Top row: Accreted-to-total gas mass fraction normalised by snapshot spacing (left) and gas velocity dispersion in the radial direction (right) plotted against the spread measure $\delta$. The points are represented as a number histogram and the red line shows the median curve. Also shown the contours enclosing a given percentage of the points. Bottom row: Total surface density and gas fraction plotted against the spread measure $\delta$. The points are represented as a number histogram, the red line shows the median curve. Also shown the contours enclosing a given percentage of the points. There is an anti-correlation in the case of total surface density and a correlation with gas fraction. These trends arise as a consequence of the radial dependence of $\delta$ and the radial dependence of these quantities and do not indicate a direct physical relation.}
\label{surfdens}

\end{figure*}

In Figure \ref{surfdens}, we plot $\delta$ against the four properties which correlate most strongly with it. The median curves are plotted above the density histograms to show the trends more clearly. Again, these plots have a non-trivial amount of scatter but also well-defined loci where we have the highest point density. We choose to present $\delta$ here, rather than $w$, as the trends seen are qualitatively similar and $\delta$ is the quantity we decide to parametrise in the following section.

We can see in Fig. \ref{surfdens} that $\delta$ increases with increasing gas fraction, increasing accreted gas fraction, and decreasing total (and gas) surface density. We also find that there is an increasing trend with the velocity dispersion $\sigma\sub{tot}$, which is mainly driven by the radial component $\sigma\sub{r}$. There is no trend seen with SFR or Toomre Q. It essential to differentiate which of these trends are just correlations with radius, and which have an independent contribution. For example, the increase in $\delta$ and decrease in $\Sigma_{tot}$ with radius naturally leads to an anti-correlation between $\delta$ and $\Sigma_{tot}$, but does not necessarily mean that the two are causally connected. 

The correlation with $\sigma\sub{r}$ can be understood on physical grounds since the tracers in a ring with high velocity dispersion, are more likely to travel to larger distances resulting in broader histograms with higher values of $\delta$. Concerning $\dot{f}\sub{acc}$, a larger amount of accreted material is likely to disturb the existing material in the ring, driving radial motions. There is a similar, although weaker, trend with the accretion rate to the ring.

\begin{figure}%[ht!]
\centering
\includegraphics[width=\linewidth]{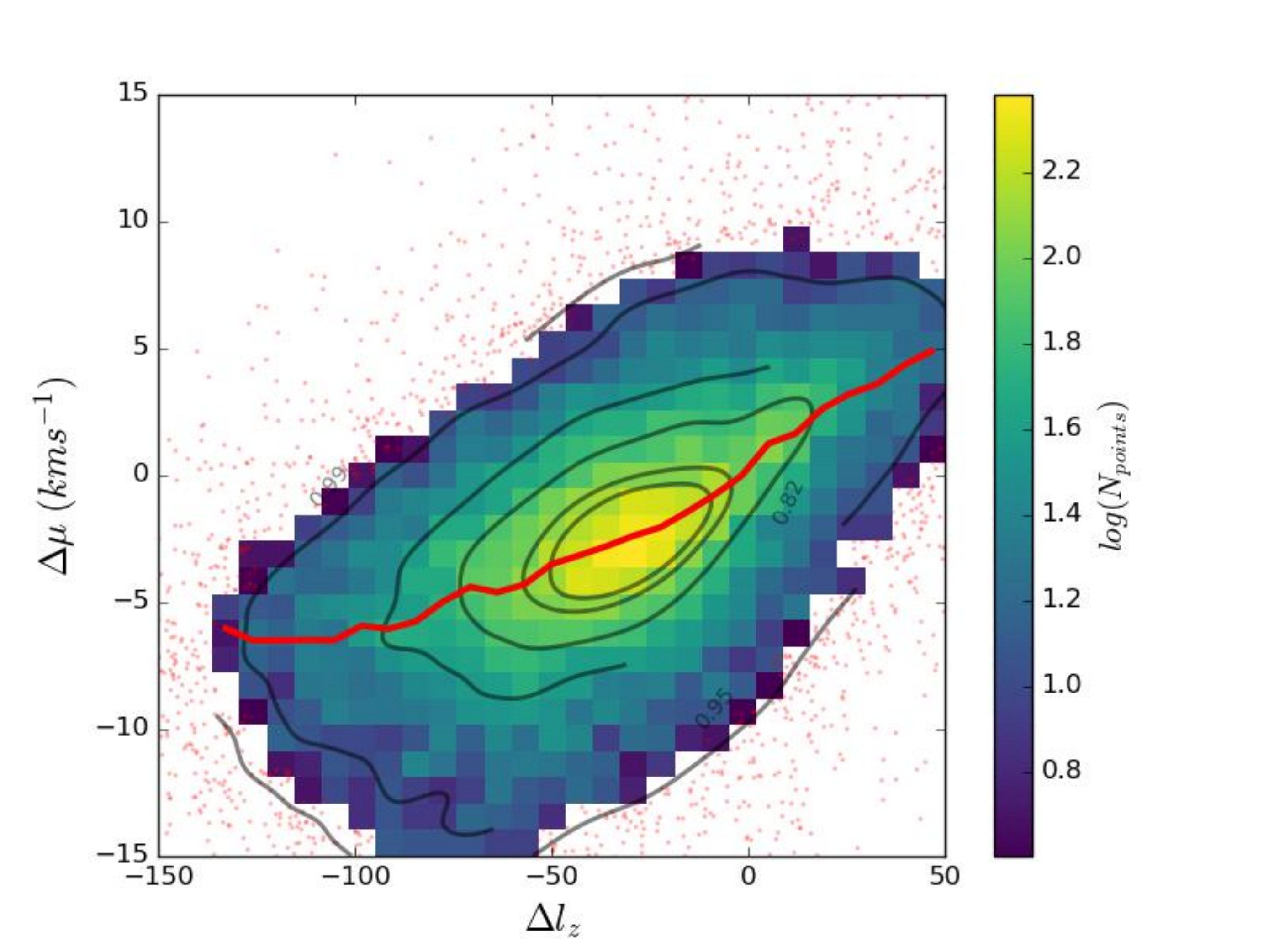}
\caption{The correlation between the median shift and the change in the specific angular momentum of the tracers is an indication that we observe a phenomenon where the inwards motion of gas (negative $\Delta \mu$ values) is associated with a loss of angular momentum (negative $\Delta l_{z}$).}
\label{medLz}
\end{figure}

Concerning $\Delta \mu$ there are only weak trends with the gas accretion rate, accreted gas fraction, and the velocity dispersion. Larger accretion and velocity dispersion lead to more negative velocities (\ie{}larger inflow speeds). The quantity that correlates most strongly with $\Delta \mu$ is the mean change in the specific angular momentum of the gas, as shown in Fig. \ref{medLz}. The specific angular momentum in the z-direction of a gas cell is expressed as $l\sub{z}=|\mathbf{x} \times \mathbf{v}|$ or simply $rv\sub{rot}$. We calculate the change in angular momentum $\Delta l\sub{z}$ for each tracer by taking the difference in the angular momentum in the $z$-direction (\ie{}out of the plane of the disc), $l\sub{z}$, between snapshots $n$ and $n+1$. The values for $l\sub{z}$ are drawn from the parent gas cell for each tracer as it is for the other tracer properties. The correlation between $\Delta \mu$ and $\Delta l\sub{z}$ is expected, since a loss of rotational angular momentum will lead to inward motions, expressed as the negative change in the gas' median position. Following the definition of $l\sub{z}$ and since most of the gas is in nearly circular orbits in the disc and the rotational velocity curves are reasonably flat, a change in angular momentum $\Delta l\sub{z}$ is correlated with a change in radius, which is expressed as the median shift, $\Delta \mu$, in our case. Further insight is needed with regard to the process that causes the angular momentum change, and hence the bulk flow, in each case.

\subsection{Identifying the strongest correlations and causations in the data}

We have tested for secondary dependencies of $\delta$ and $\Delta \mu$ at fixed radius by plotting the residuals around the median $\delta-r$ and $\Delta \mu - r$ relations. The residual is simply the distance of a given data point from a fit to the median relation, which in the case of $\delta-r$ is parameterised as a power law and in the case of $\Delta \mu - r$ as a piece-wise linear fit. Looking at the residuals allows us to make a distinction between quantities that are actual drivers of trends in $\delta$ and $\Delta \mu$, and those that only correlate because of a third property (in our case the radius). We quantify the strength of the relation between the residual and a secondary property by calculating the correlation coefficient between the two. Table \ref{tabRes} shows the values of these correlation coefficients for the selected quantities, both for the residuals in $\delta$ and $\Delta \mu$. A higher absolute value of the correlation coefficient is an indication that this quantity is more likely to drive the scatter we observe around the median. 

\begin{figure}
\centering
\begin{minipage}{.5\textwidth}
\includegraphics[width=\linewidth]{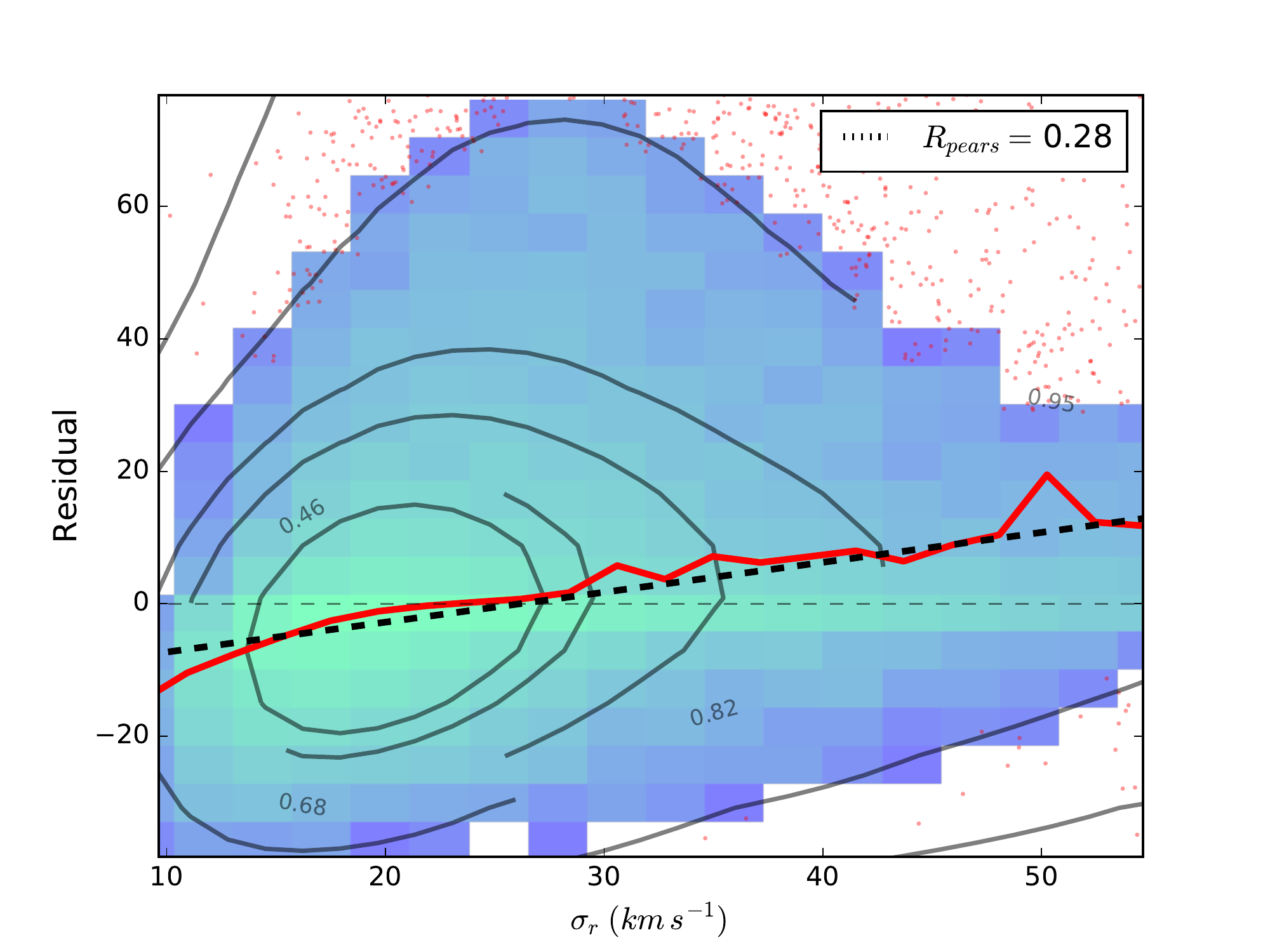}
\end{minipage}
\begin{minipage}{.5\textwidth}
\includegraphics[width=\linewidth]{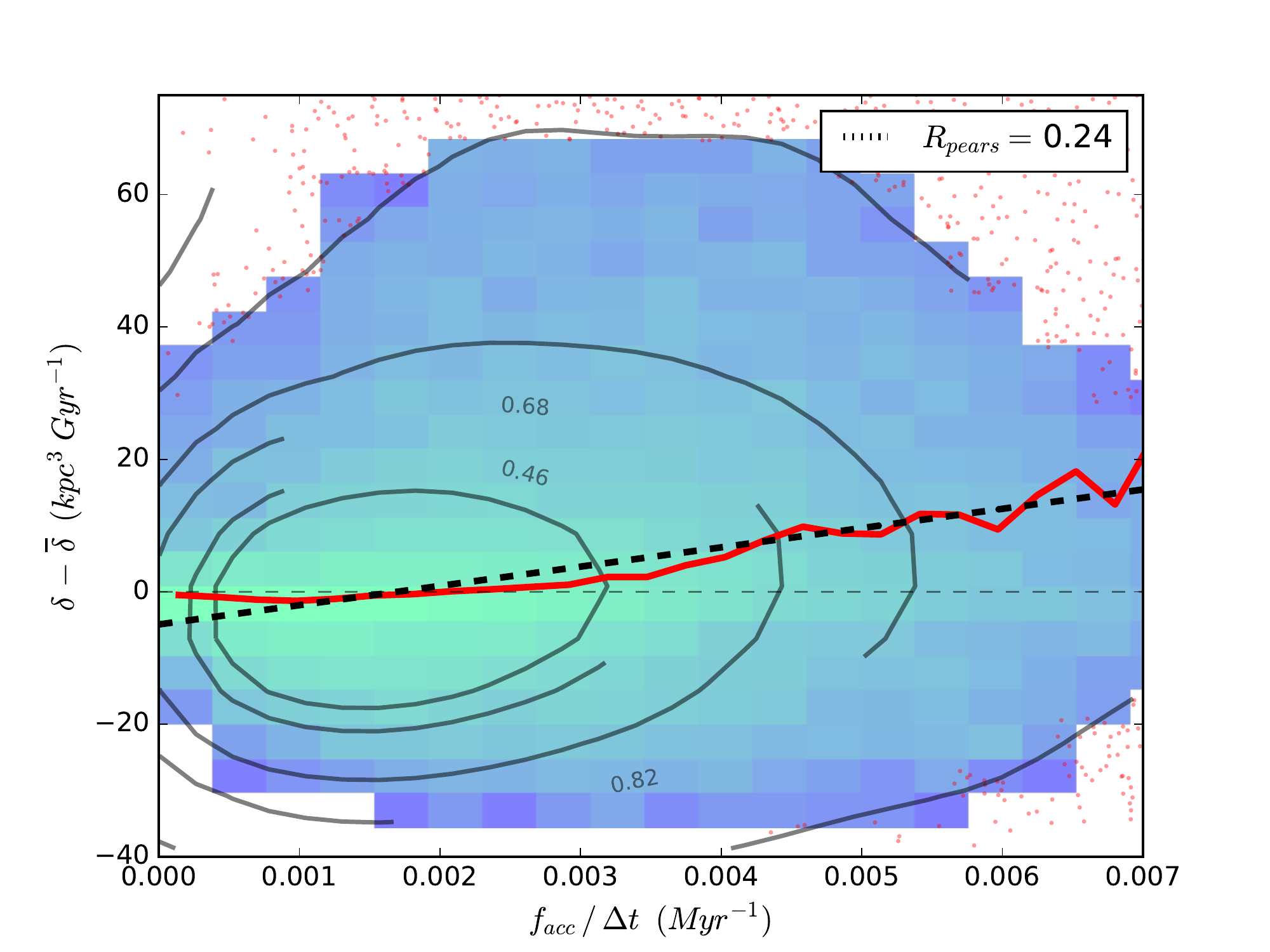}
\end{minipage}
\caption{The residuals in the around the median in the $\delta-r$ plot correlate with the radial velocity dispersion of the gas. At a given radius, higher velocity dispersion of the material leads to larger values of the spread $\delta$. Positive values for the residuals mean that at the given ring the measured spread is above the median curve of the whole sample.}
\label{residuals_w}
\end{figure}

First of all, we find that the residuals do not show evidence of time dependence as there is an absence of correlation with redshift, and nor any correlation with a specific halo. $f\sub{gas}$ is an example of a quantity that shows positive correlation with $\delta$ but no trend with the $\delta-r$ residuals. On the other hand, the velocity dispersion $\sigma\sub{tot}$ has a positive correlation with the $\delta-r$ residuals. Upon splitting the velocity dispersion into different components, we find that this correlation is driven mostly by the dispersion in the radial direction $\sigma_{r}$. In other words, the scatter in the $\delta-r$ plane is produced primarily by the different $\sigma\sub{r}$ among rings at a given radius. Differences in the accreted gas fraction also play a role in producing the scatter seen. The residuals as a function of $\sigma\sub{r}$ and $\dot{f}\sub{acc}$ are shown in Fig. \ref{residuals_w}. We present the residual plots of the quantities that correlate more strongly.

\begin{figure}
\centering
\includegraphics[width=\linewidth]{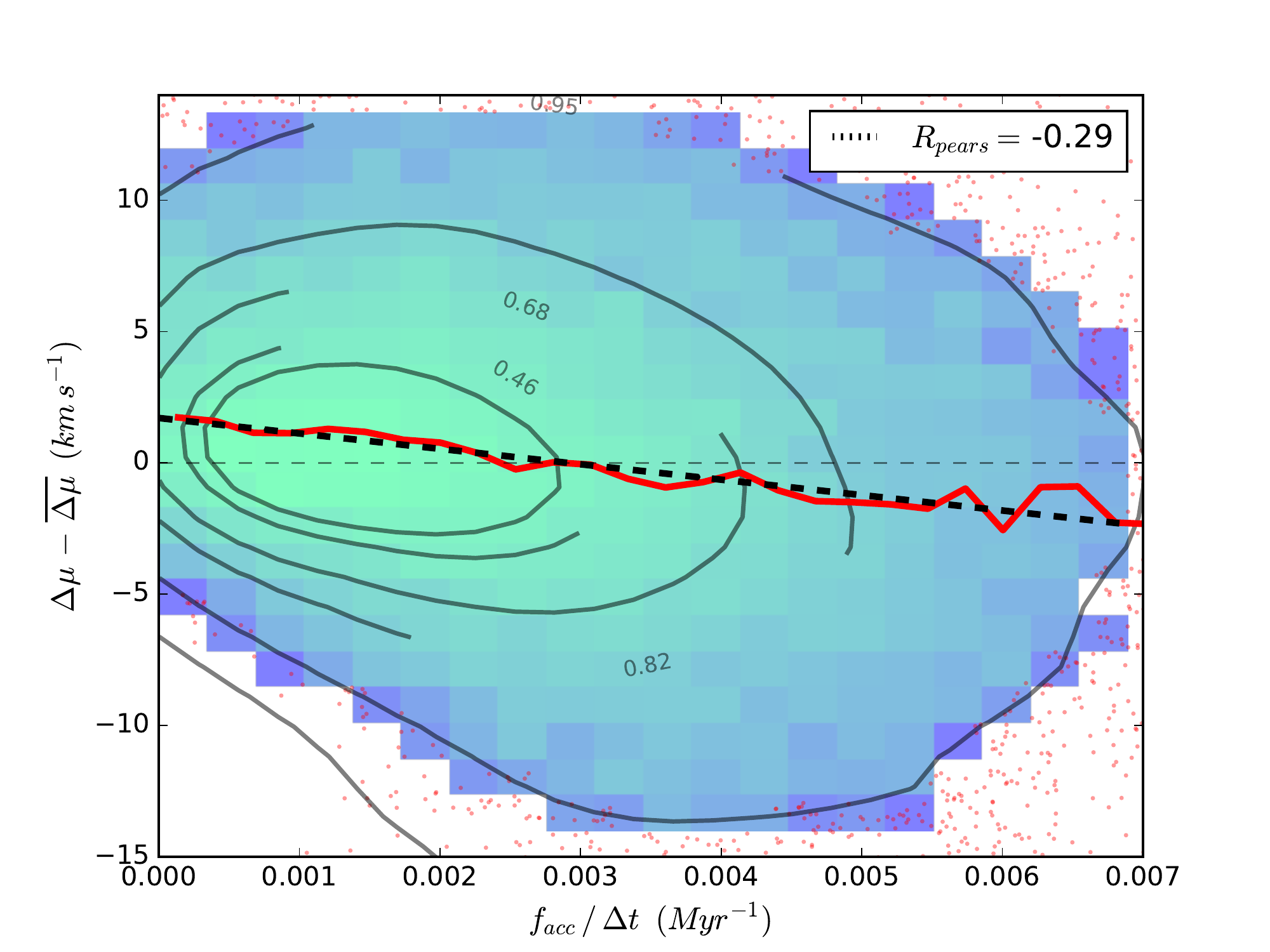}
\caption{There is a weak correlation of the residuals around the mean $\Delta \mu$ with the accreted gas fraction ($\dot{f}_{acc}$), where the more gas is accreted compared to the existing gas in the ring the larger is the inflow speed.}
\label{residuals_dm}
\end{figure}

Regarding $\Delta{}\mu$, Table \ref{tabRes} shows that there is a weak but clear anti-correlation with $\dot{f}_{\rm acc}$, followed by a positive correlation with the surface density. This is reasonable, since the primary quantity from which we extract the residuals is the radius, and since $\Delta \mu$ shows no correlation with radius in the disc proper, the direct relation of it with $\dot{f}\sub{\rm acc}$ is reflected in the residuals. The residual plot for $\Delta \mu$ as a function of $\dot{f}\sub{\rm acc}$ is shown in Fig. \ref{residuals_dm}.  

Based on the information from the residuals discussed above, we include the quantities with the strongest residual correlations alongside radius in the final parameterisation.

\begin{table}
\centering
 \begin{tabular}{|c|c|c|} 
 \hline
 Quantity & $R\sub{corr}$ with $\delta$ & $R\sub{corr}$ with $\Delta \mu$\\ with $w$\\
 \hline
 $z$ & -0.09 & 0.07\\
 $\Sigma\sub{tot}$ & -0.01 & 0.18\\
 %\hline
 $\Sigma\sub{gas}$ & 0.06 & 0.16\\
 %\hline
 $f\sub{gas}$ & 0.09 & -0.07\\
 %\hline
 $\sigma\sub{tot}$ & 0.25 & 0.08 \\
 %\hline
 $\sigma\sub{r}$ & 0.28 & 0.03\\
 %\hline
 $\sigma\sub{z}$ & 0.23 & -0.04\\
 %\hline
 $\dot{M}\sub{acc}$ & 0.11 &  0.22\\
 %\hline
 $f\sub{acc}/\Delta t$ & 0.24 & -0.29\\
 %\hline
 SFR & -0.11 & 0.14 \\
 \hline
\end{tabular}
\caption{Evaluation of the correlation factor, $R\sub{corr}$, between the residuals and the quantities of interest. A stronger correlation coefficient is an indication that the particular quantity is more important in influencing the scatter in the $\delta-r$ or $\Delta \mu-r$ plots.}
\label{tabRes}
\end{table}

\subsection{Best fits}

The mean evolution of $\delta$ with radius can be fit accurately with a power law of $\sim 1.1$ (Fig. \ref{delta_rdepend}). The power law fit is slightly preferred over a linear fit in $r$ because it better describes the $\delta-r$ dependence in the innermost parts of the discs. Of all the secondary quantities that we consider, $\dot{f}\sub{acc}$ and $\sigma\sub{r}$ show the strongest correlations in the residuals around the mean $\delta-r$ curve (see Table \ref{tabRes}). We normalise the secondary quantities with some characteristic values to always have non-dimensional terms in the right-hand side of the parametrisations.

Our final parameterisation is the combination of the power law fit to the radius and a linear fit to the secondary quantity, extracted from the residual information. Consequently, we present two possible parametrisations:

\begin{align}
    \delta\ /\ \tn{kpc}^{3}\,\tn{Gyr}^{-1} = & \, 35.9\,(r/R_{95})^{1.1} + 21.8\,\left(f_{\rm acc}*\frac{60 \rm Myr}{\Delta t}\right) - 2.8\label{wfit1}\\
    \nonumber & \\
    \delta\ /\ \tn{kpc}^{3}\,\tn{Gyr}^{-1} = & \, 35.9\,(r/R_{95})^{1.1} + 14.9\,\left(\frac{\sigma\sub{r}}{40\,\tn{km s}^{-1}}\right) - 9.9 \label{wfit2}
\end{align}

\begin{figure*}
\centering
\begin{minipage}{.5\textwidth}
  \centering
  \includegraphics[width=\linewidth]{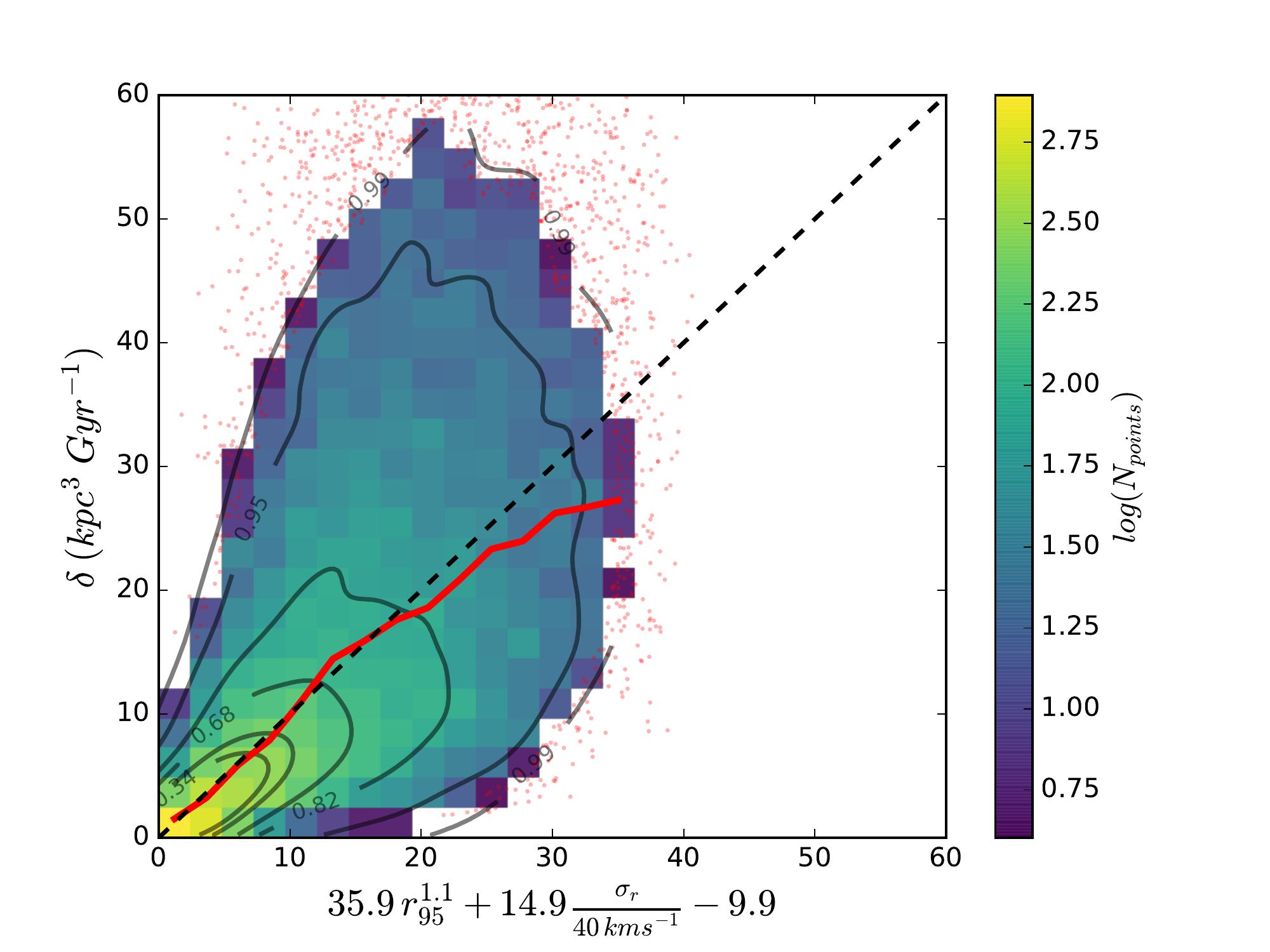}
\end{minipage}%
\begin{minipage}{.5\textwidth}
  \centering
  \includegraphics[width=\linewidth]{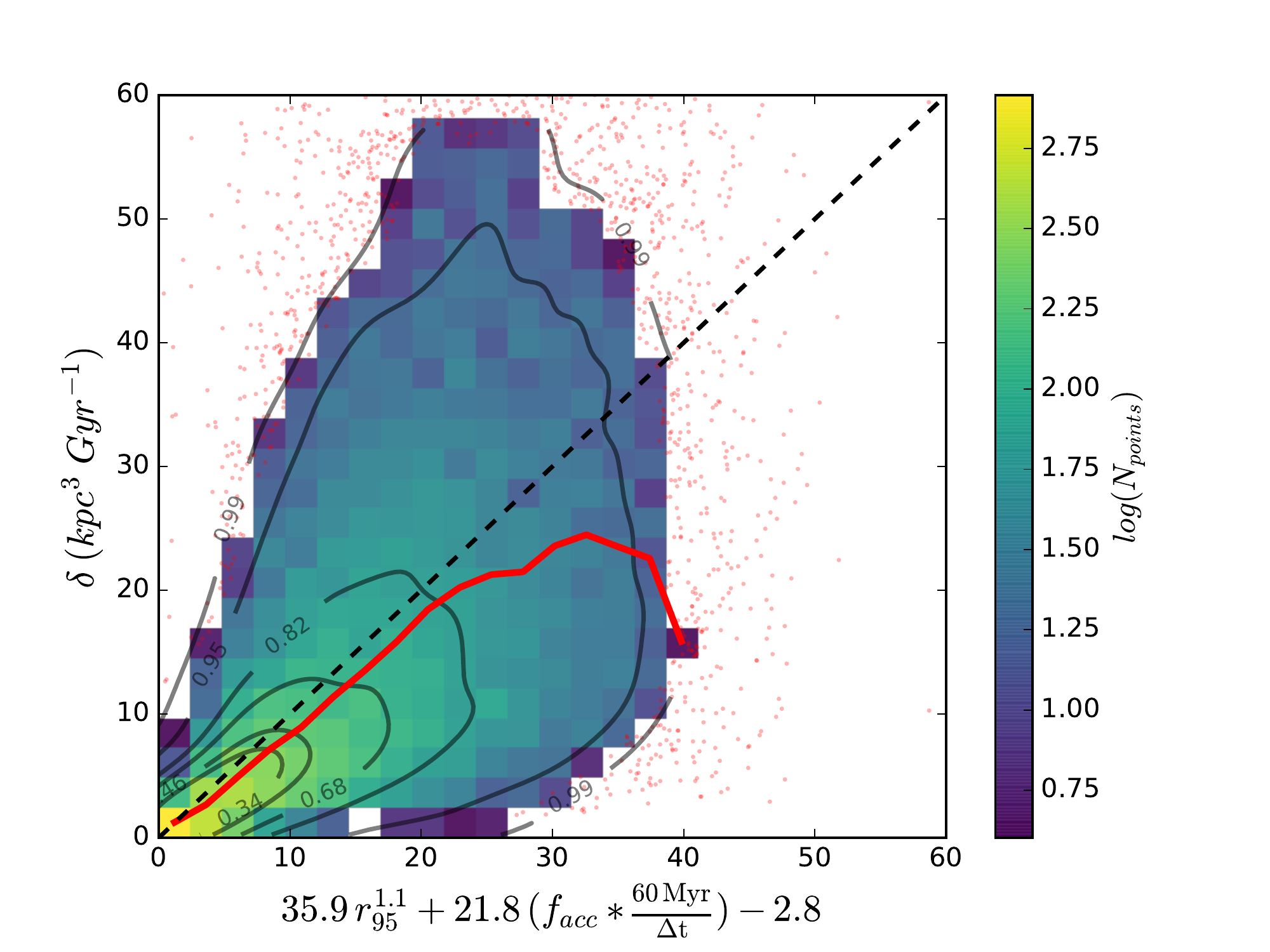}
\end{minipage}
\caption{The selected parametrisations for the quantity w, using two different sets of parameters, radius - accreted gas fraction (left) and radius - radial velocity dispersion (right). In these plots the data should be lying around the one-to-one relation if the fitting is ideal. This line is plotted for reference (dashed). The red line is the median of the data which for the most part agrees well with the dashed line.}
\label{lincombs}
\end{figure*}

In Fig. \ref{lincombs}, we show the calculated $\delta$ using these parametrisations and plot it against the actual value for $\delta$ for each ring measured from the data. The median line for the dataset in this $\delta_{\rm meas}-\delta_{\rm calc}$ plot lies on the 1-1 relation (dashed black line) out to around $20\, \tn{kpc}^{3}\,\tn{Gyr}^{-1}$, as expected. The scatter around the 1-1 relation follows from the scatter around the linear fit of the residual plots. 

There is a set of points for $\delta \gtrsim{}30\,\tn{kpc}^{3}\,\tn{Gyr}^{-1}$ (or equivalently $w\gtrsim{}1.6$) that are not well-described by the parameterisation. This is a consequence of how the surface created by the parameterisation traces the 3D point distribution of $r-f\sub{acc}-\delta$ or $r-\sigma\sub{r}-\delta$. Isolating these points and trying to identify if they are caused by some specific process or depend on a given property shows no conclusive results. This is not a big concern, as these points account for less than 20 per cent of the data. They are found mostly in the outer parts of the discs and may be caused by residual merger interactions but also gas accretion.

With regards to a parameterisation for $\Delta \mu$, we can fit the inner part of the disc ($r<0.75*R_{95}$) with a constant with respect to radius, which from the data is found to be $-2.4\,\tn{km s}^{-1}$ and the outer part ($r>0.75*R_{95}$) with a  linear fit indicating faster inflow speed. The value of 0.75 is found by applying the fit. The scatter around the fit is then given by the residual plots of either $\Delta l_{z}$ or $\dot{f}_{\rm acc}$. However, $\Delta l_{z}$, as mentioned before, is merely a different expression of $\Delta \mu$ in the case of a flat rotation curve, so it is not very informative to build a parameterisation of $\Delta \mu$ in terms of it. $\dot{f}_{\rm acc}$ can be used as a secondary parameter as it is an independently measured quantity of an external process that could potentially be a driver of the bulk flows.

For the purposes of arriving at a parameterisation that can be useful in semi-analytic models, we thus arrive to the following equations:

\begin{align} \label{mufit1}
\overline{\Delta \mu}\,/\,\tn{km s}^{-1} = \bigg{\{} &         
    \begin{array}{ll}
		-2.4 & \tn{if } r \leq 0.75\,R_{95} \\	
		-15.9\,(r/R_{95}) + 9.5 & \tn{if } r > 0.75\,R_{95}
	\end{array}% \;\;,
\end{align}
If we further include the parameter $\dot{f}\sub{acc}$ to describe the scatter alongside the median relation, the above equations are modified to

% \begin{align} \label{mufit2}
% \Delta \mu\,/\,\tn{km s}^{-1} = \bigg{\{} &         \begin{array}{ll}
% 		-1.7 - 6.8\,(f_{\rm acc}*\frac{60 \rm Myr}{\Delta t}) \;\; \tn{if } r \leq 0.75\,R_{95} & \\
% 		-15.9\,(r/R_{95}) - 6.8\, (f_{\rm acc}*\frac{60 \rm Myr}{\Delta t}) + 10.2  &\\
% 		\tn{if } r > 0.75\,R_{95}& \\
% 	\end{array}% \;\;,
% \end{align}

\begin{align} \label{mufit2}
\nonumber \Delta \mu & \,/\, \tn{km s}^{-1} = \\
& = \bigg{\{} \begin{array}{ll}
		-1.7 - 6.8\,(f_{\rm acc}\,\frac{60 \rm Myr}{\Delta t}) \ \ \ \ \ \ \ \ \ \ \ \ \ \ \ \ \ \ \ \ \tn{if } r \leq 0.75\,R_{95} & \\
		-15.9\,\left(\frac{r}{R_{95}}\right) - 6.8\, (f_{\rm acc}\,\frac{60 \rm Myr}{\Delta t}) + 10.2\ \ \tn{if } r > 0.75\,R_{95}& \\
	\end{array}% \;\;,
\end{align}

% \begin{align} \label{mufit2}
% \nonumber \Delta \mu\,/\, & \tn{km s}^{-1} = \\
% \nonumber & = -1.7 - 6.8\,(f_{\rm acc}*\frac{60 \rm Myr}{\Delta t}) \;\; & \tn{if } r \leq 0.75\,R_{95} \\
% & = -15.9\,(r/R_{95}) - 6.8\, (f_{\rm acc}*\frac{60 \rm Myr}{\Delta t}) + 10.2  & \tn{if } r > 0.75\,R_{95}& \\
% \end{align}

%\begin{align} \label{mufit2}
%\Delta \mu\,/\,\tn{km s}^{-1} = \bigg{\{} &        
%    \begin{array}{ll}
%        -1.7 - 6.8\,(f_{\rm acc}*\frac{60 \rm Myr}{\Delta t}) & \\
%        \hskip10em\relax \tn{if } r \leq 0.75\,R_{95} & \\	
%		-15.9\,(r/R_{95}) - 6.8\, (f_{\rm acc}*\frac{60 \rm Myr}{\Delta t}) + 10.2 & \\
%		\hskip10em\relax \tn{if } r > 0.75\,R_{95}
%	\end{array}% \;\;,
%\end{align}

These parameterisations give a most accurate description in the regime of values $-10<\Delta \mu<0$, which contain the majority of points, but are not representative for cases with $\Delta \mu > 0$, where we have radial outflow of the material.

\section{Discussion} \label{sec:Discussion}

We have identified the accreted gas fraction, $\dot{f}\sub{acc}$, and gas velocity dispersion, $\sigma$, as the two main parameters driving variations in gas spread, $w$ (or its timestep-invariant equivalent, $\delta$), with radius in the Auriga simulations. On physical grounds, $\sigma$ in a given ring is partially a measure of the total internal kinetic energy and the amount of turbulence that is present in the gas. This is the case no matter which mechanism injected the energy into the system, be it for example kinetic heating from some interaction or stellar feedback. Furthermore, as we study radial motions, the radial component of the dispersion, $\sigma\sub{r}$, is expected to be more dominant. Studies like \cite{Forbes14} and \cite{Yang12}, modelling the diffusion of metals in the disk, suggest diffusion coefficients scaling with the velocity dispersion of the gas multiplied with the scale height of the disk. We have tested whether such a quantity ($\sigma_{g} h_{g}$) shows any relation with the spread measure $\delta$ and we find that it yields similar strength of correlation to the residuals to when simply using $\sigma_{g}$ as a parameter.

The accretion process is also very relevant to the radial motions, as has also been shown in earlier studies \citep{Pezzulli16}. The accretion rate of new gas could also be a candidate parameter but we found that $\dot{f}\sub{acc}$ correlates better with $w$ and $\delta$. Besides, $\dot{f}_{\rm acc}$ carries more information, as it is a measure of both the material entering the ring and the material already present. 
We could connect the effect of the accreted gas to the radial motions by considering that larger amounts of accreted material result in more kinetic energy that can be converted into turbulence, leading to larger random radial motions which then translate to the larger values of $\delta$. Especially at the outer, lower-density regions of the disc, turbulence can dominate the energy density, as low-density gas has less inertia and responds more readily to perturbations from the external material.

The fact that w scales as $w \sim \Delta t^3$ is not straightforward to justify. In a simple diffusive process, where gas diffuses  out of the initial ring to lower density regions, we would expect a $w \sim \Delta t^2$ dependence. The cubic power that we find instead gives a better fit, suggestive of a process slower than pure diffusion. The overall radial spread of the gas in the disc is likely the result of a combination of physical processes, some of which are of diffusive nature, that are active during the disc evolution within the disc plane. On top of this, it must be noted that the cubic power is the average of all the data in the 14 different halos and over the whole redshift range that we use. Thus, we cannot clearly state why $w \sim \Delta t^3$, but only acknowledge that this time dependence better brings the data from different snapshot spacings in agreement.

\begin{figure}
\centering
\includegraphics[width=\linewidth]{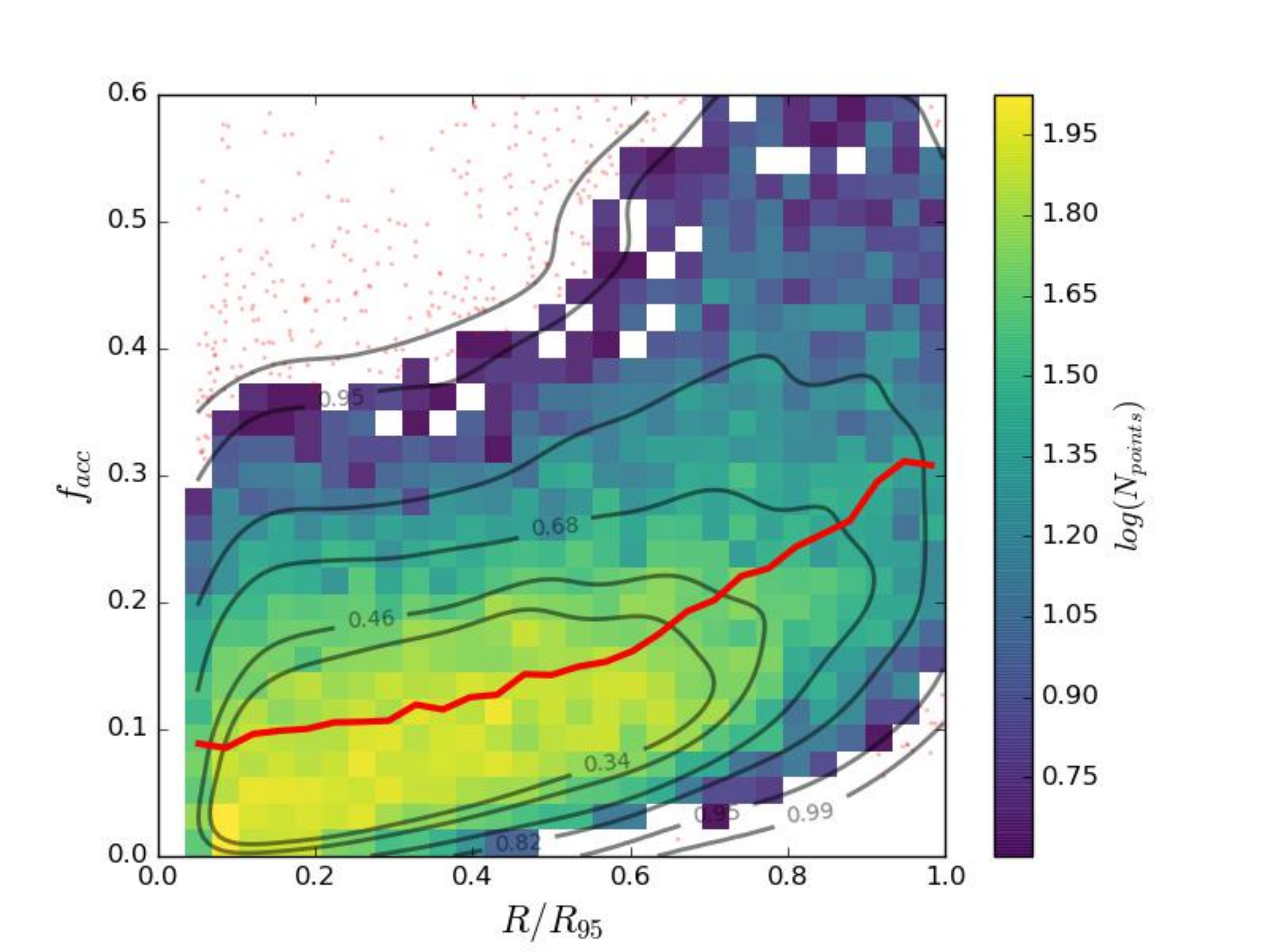}
\caption{Radial dependence of the accreted gas fraction. At the outer parts of the discs, beyond 70\% of the disc radius, we have larger contribution from the accreted material.}
\label{medFacc}
\end{figure}

The radial dependence of $\Delta \mu$, the bulk flow, as seen in Fig. \ref{radmed}, can be explained by two separate regimes in the disc. The regime where we observe a nearly constant radial dependence with inflow a few $\rm km\,s^{-1}$, and the one where there is inflow with much larger velocities, increasing as we head in the outer parts of disc. In the first case, we are essentially describing the equilibrium part of the disc where the material has settled in more regular motions and is rotationally supported. In the second case, we are in a regime where we have significant accretion of new material, coming in as patchy accretion in many cases. These blobs of gas can travel relatively unimpeded until they encounter the comparatively higher densities at the edge of the star forming disc. Fig. \ref{medFacc} shows evidence for this statement, as beyond $\sim 0.75R_{95}$ we find higher $f\sub{acc}$ values and a steeper slope in its radial profile. \cite{Goldbaum15b} have calculated the time-averaged radial gas mass flux in their simulated galaxy, finding a radial profile that points to a net inflow with little radial evolution in the absolute value of the flux, which can be consistent with the radial profile that we find for $\Delta \mu$. Aside from these two regimes, we also attribute the surplus of positive (outflowing) $\Delta \mu$ values at $r\lesssim0.1$ to AGN feedback, which empties of gas the innermost regions of galaxies with active black holes. This effect appears strongly only in 3 halos in the sample, for specific timespans, so does not significantly alter our conclusions. We must also notice that the constant value that we find in the inner parts for the inflow is representative of the set of the halos that are available in Auriga and is likely limited to the specific mass range. We have no indication that is a value that can be generalised to very different galaxy mass regimes.

The range of values that we find for $\Delta \mu$ are consistent with the observational data from \cite{Schmidt16} where they find that most of their data points are within $\pm15 \; \textrm{km\, s}^{-1}$. The exact radial flow speed profiles presented in this paper may not necessarily match the average radial profile we show in Fig. \ref{radmed} but this is expected as we present the compilation of all the data for a large number of snapshots. The galaxies used in \cite{Schmidt16} show an object-to-object variability with strong inflows or outflows at given objects exceeding $\pm30 \; \textrm{km\, s}^{-1}$ and at different radii. This is also true in our simulations if we look at specific snapshots where we have instances of extreme inflows or outflows, comparing to the average, with no clear radial trend. Concerning the values of the spread $w$, it is much harder to test against observations since it is not a directly measurable quantity in observational data.

The $\Delta \mu$ parametrisation we provide describes the average behaviour we observe over all halos in the suite. As seen from the data, there are many instances where tracers move on average outwards between two snapshots. This is not captured in the best fit, which gives only a time-averaged representation. The correlation with $\Delta l_{z}$ shows us that the gas moves inwards or outwards because its angular momentum has been altered. This indicates the presence of a torque that has driven this loss or gain. However, identifying the source of this torque, and more importantly reliably connecting it with the movement of gas, is a difficult proposition. One possibility is the presence of spiral arms that by interacting with the gas can input or remove angular momentum from it. 

Both \cite{Krumholz18} and \cite{Goldbaum15b} discuss the relevance of the Toomre Q parameter in radial flows, and although we have examined the Q values for our model discs, we found no convincing dependence between them and $\Delta \mu$ or $\delta$ but only weak correlations with a lot of scatter and driven by high values of Q. Given that in the \cite{Krumholz18} model and its variants, Q is often set to a constant value, or subject to a floor value, we should not necessarily expect a correlation but the lack of it means that we cannot use Q in the way we construct the parametrisations. All in all, we do not rule out the importance of gravitational instability as a source of turbulence, but rather suggest that the ring analysis we perform may not capture this effect. Further, the ISM model in the Auriga simulations, which is designed to prevent clump formation and generally yields higher Q values, is not conducive in resolving perturbations from gas clumps that could be a main physical reason underlying any dependence on Q. 

We have tested the resolution dependence on one of our halos that was re-simulated with lower resolution, to evaluate the consistency in the results that we obtain. The results between the two resolution simulations of this single halo are mostly consistent within the error, but the lower resolution simulation shows overall higher values for $w$ (on average 1.3 times higher) with the effect being more pronounced in the very inner radii where also $\Delta \mu$ appears to deviate from the fiducial run. In other words, at lower resolution, with a lower number of tracers (similar number of tracers per cell but lower amount of cells overall), the tracers appear more diffusive. In general, gas flows are less well captured in the lower resolution simulation because of the low number of tracers that sample the cold gas.

Further it must be noted that by using ring-like annuli in our analysis we smooth out any azimuthal variation in the two quantities we study. For example, the presence of a strong bar can lead to material funneling to the centre at particular azimuthal angles but being expelled in another direction. This information in a given ring is captured in the spread $\delta$, resulting in a symmetric distribution but the median bulk flow $\Delta \mu$, being the average value of the speeds of inflowing and outflowing material, will be lower than if we look at the speed of material in a particular direction.

We see a small difference in the merger history between the 6 higher-mass ($1-2 \times 10^{12} M_{\odot}$) and the 8 lower-mass ($0.5-1 \times 10^{12} M_{\odot}$) halos. In almost half of the lower-mass haloes, there are mergers and encounters even at later stages, whereas the higher-mass ones are relatively quiet. This may indicate that the higher-mass sample is in a slightly different evolutionary stage, but this does not seem to influence the conclusions for the properties that describe the radial flows.

As a final remark, we acknowledge that the simulations do not explicitly model the small-scale turbulence generated by stellar feedback and could impact the radial movement of the tracers on small scales, but that the effective pressure applied by the sub grid model provides some similar effect to the turbulent pressure in star-forming gas. Getting a better understanding of these effects would require simulations that explicitly model the multi-phase ISM, which is beyond the scope of this paper.

\section{Conclusions} \label{sec:Conclusions}

We have performed an analysis of the gas kinematics in disc galaxies in the Auriga simulation suite. We have focused only in the `quiet' phases of the disc evolution, excluding the snapshots when the discs have a violent merger. In our method, we examine the disc in a local fashion, by considering a ring of gas at a given radius. We describe the radial flows of gas with two parameters; the median bulk flow, $\Delta{}\mu$, and radial spread, $w$, of the gas in each ring. We have identified $\delta=w^{3}/\Delta   t$ as a timestep invariant quantity. As the radius increases, we observe an increase in $w$ and hence $\delta$, indicating that tracers in the outer regions diffuse out of the initial ring more effectively than in the inner regions. This can be attributed to the lower densities (of gas and stars) or the larger accretion rates observed at larger radii. The bulk flows expressed by $\Delta \mu$ have a flat radial dependence in the inner parts of the disc, whereas in the outer parts we observe increased inflow speeds. Both quantities appear to be closely connected to the amount of accreted material in the disc, as expressed by the accreted mass fraction $\dot{f}\sub{acc}$. We have presented parameterisations of $\Delta \mu$ as a function of radius r and $\dot{f}\sub{acc}$, differentiating between the inner disc (equilibrium region) and outer disc (accretion-dominant region). For $\delta$, our parametrisations are expressed as functions of radius and a secondary parameter which is either $\dot{f}\sub{acc}$ or the radial velocity dispersion of the gas $\sigma\sub{r}$. In combination, these two quantities describe the process of gas mass exchange in different radii inside discs. Since we have not yet tested how the results of this study apply to models, we choose to present several different parametrisations that arise from our data, with a goal of checking their performance in a future study.

\section*{Acknowledgements}
We thank the anonymous referee for the very constructive report which helped in improving this manuscript. Part of this research was carried out on the High Performance Computing resources at the Max Planck Comput-ing and Data Facility (MPCDF) in Garching operated by the MaxPlanck Society (MPG).

\section*{Data Availability}
The data underlying this article will be shared on reasonable request to the corresponding author.

\section*{Appendix}
\renewcommand\thefigure{A\arabic{figure}} 
\setcounter{figure}{0} 

\begin{figure}
    \centering
    \includegraphics[width=\linewidth]{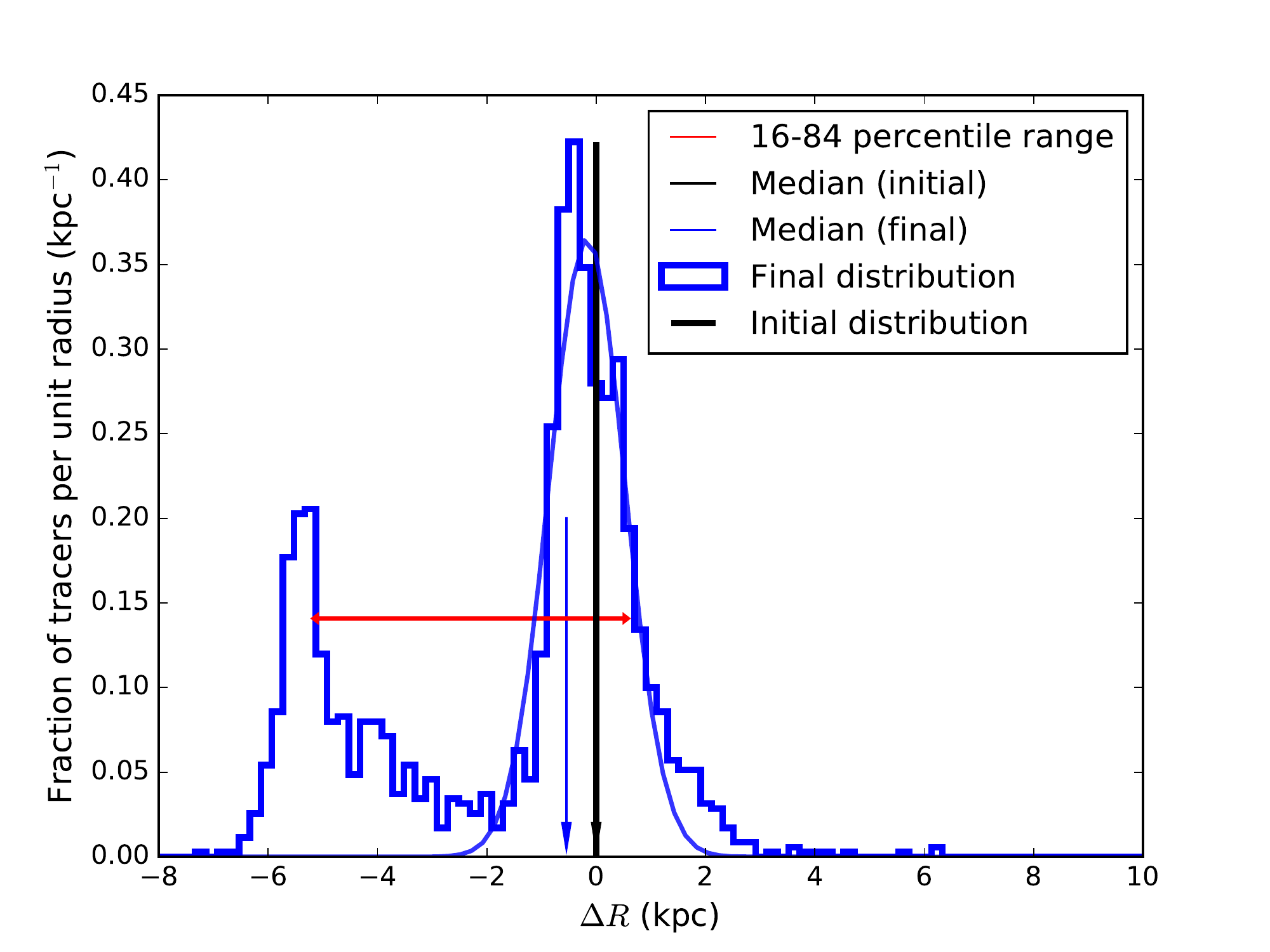}
    \caption{Asymmetric histogram example where a Gaussian is not well fit.}
    \label{badhist}
\end{figure}

\begin{figure}
	\includegraphics[width=\linewidth]{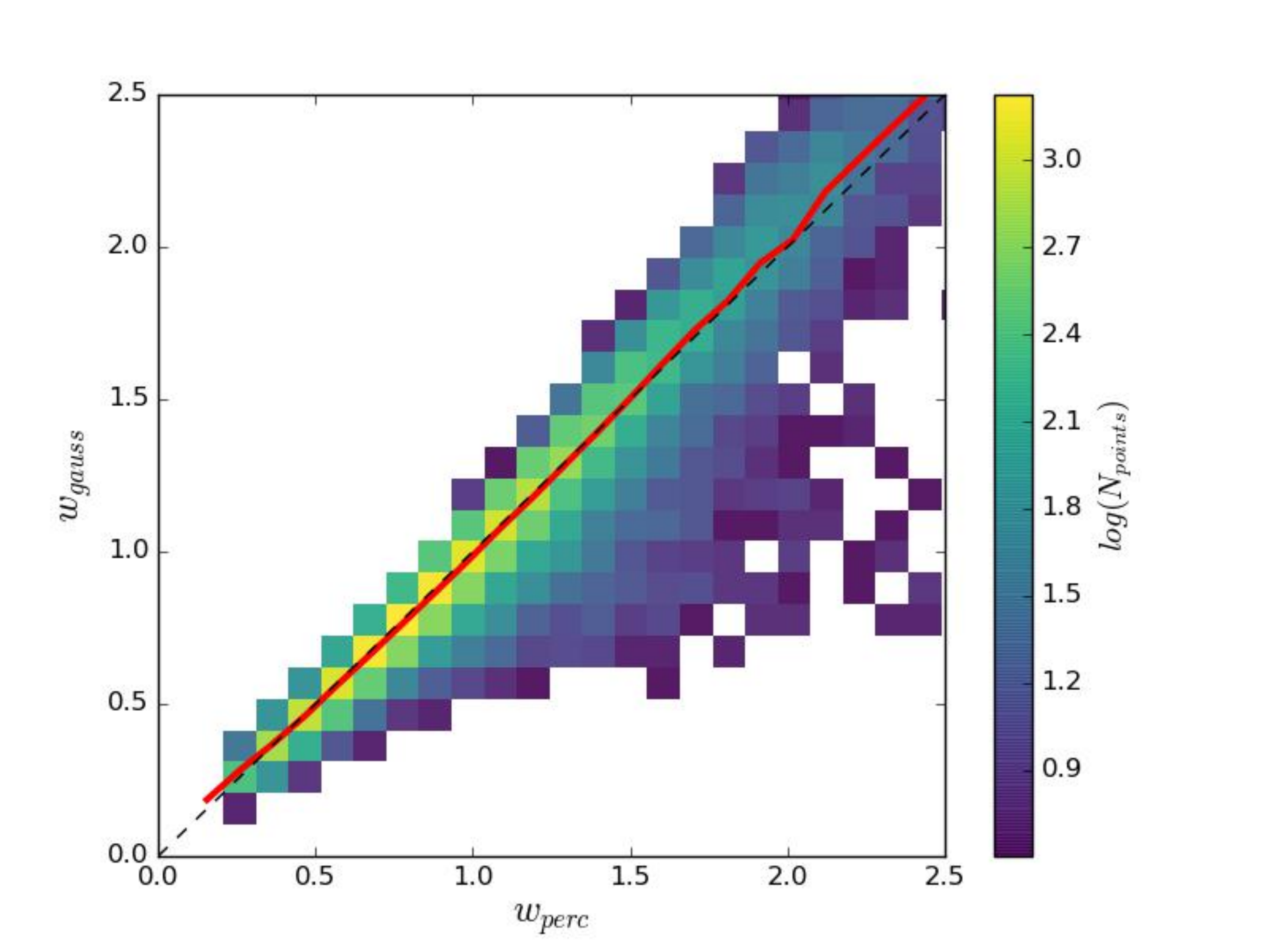}
    \caption{Comparison of the Gaussian width (y-axis) and the 16-84 percentile range values (x-axis). We observe that there is very close 1-1 correspondence of the two measurements and can mostly be used interchangeably.}
    \label{perc_gauss}
\end{figure}

\begin{figure}
	\includegraphics[width=\linewidth]{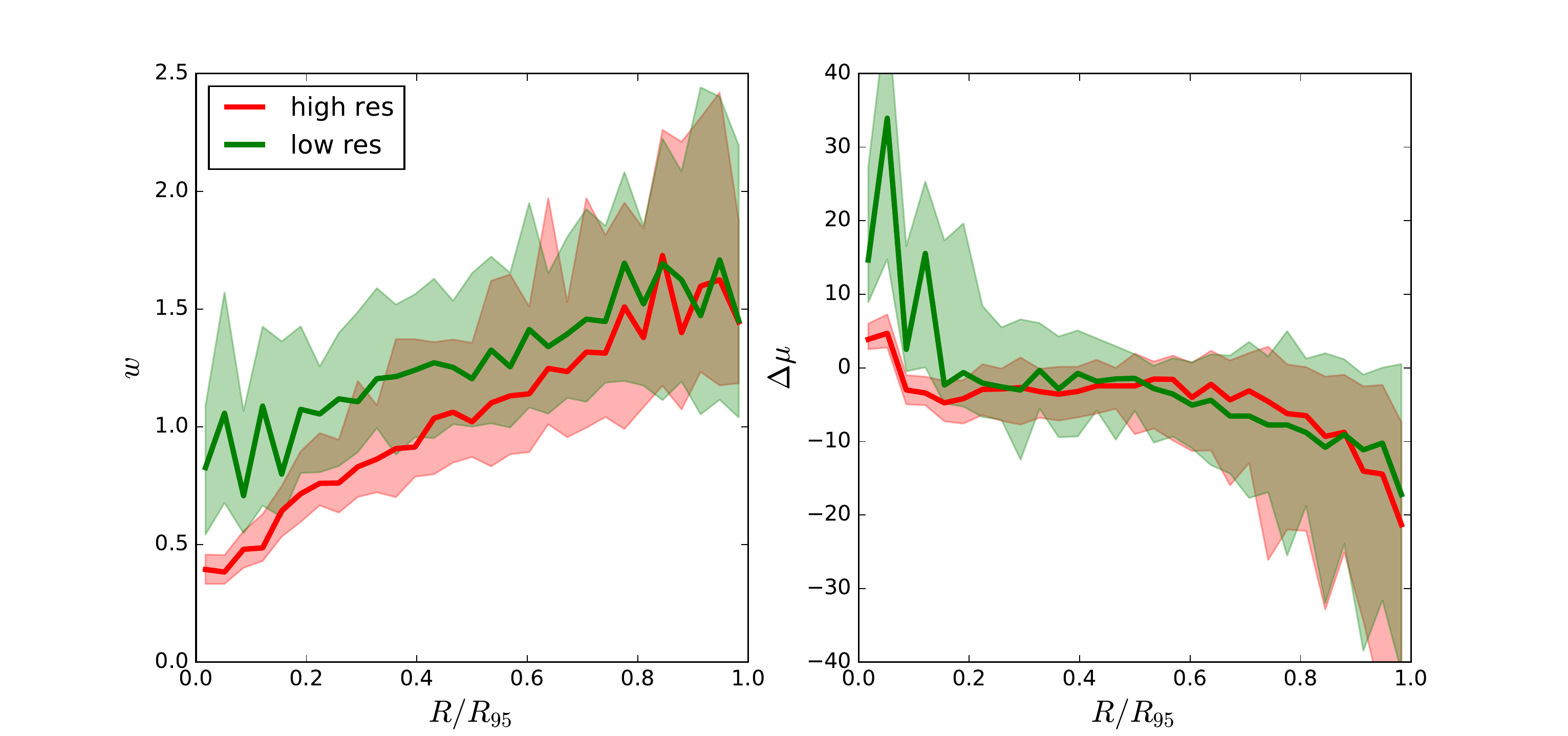}
    \caption{Comparison of the mean radial profiles of the quantities w and $\Delta mu$ when using a lower resolution simulation. This graph is made only for one halo of the set that was available in a low resolution run. The deviation from the fiducial resolution is more evident in the inner regions where tracers appear to be more diffusive. }
    \label{res_compar}
\end{figure}

Fig. \ref{badhist} shows a case where the radial distribution of the tracers at snapshot $n+1$ is highly asymmetric and a Gaussian fit is not accurately describing the shape of it. There is a considerable difference in the value of the 16-84 percentile range and the width of the Gaussian fit. This histograms appear mostly at outer regions of the discs and are probably pointing to material in the accretion phase.
In Fig. \ref{perc_gauss} we see that for the total sample of the rings the calculation of the width of the distribution described by the 16-84 percentile range and the $\sigma$ of the Gaussian fit is on average consistent. There are outlier points mostly in the lower right part of the plot which indicates that for these rings the Gaussian fit underestimates the width comparing to the percentile range calculation (as shown in Fig. \ref{badhist}). In Fig. \ref{res_compar} we examine the resolution convergence by calculating the median profiles for w and $\Delta \mu$ for a single halo from the simulation suite simulated with the fiducial and lower resolution.

%\bibliographystyle{mnras}

%%%%%%%%%%%%%%%%%%%%%%%%%%%%%%%%%%%%%%%%%%%%%%%%%%

\bibliographystyle{mnras}
\bibliography{radial_flows_paper}

% Don't change these lines
\bsp	% typesetting comment
\label{lastpage}
\end{document}